\documentclass[conference]{IEEEtran}

\newif\ifanonymous
\anonymousfalse

\usepackage{cite}
\usepackage{amsmath,amsfonts,amssymb}
\usepackage{booktabs}
\usepackage{array}
\usepackage{textcomp}
\usepackage{stfloats}
\usepackage{verbatim}
\usepackage{balance}
\usepackage{url}
\usepackage{xspace}
\usepackage{graphicx}
\usepackage{multirow}
\usepackage{tcolorbox}
\tcbuselibrary{skins, breakable}
\usepackage{algorithm}
\usepackage{algpseudocode}
\def\BibTeX{{\rm B\kern-.05em{\sc i\kern-.025em b}\kern-.08em
    T\kern-.1667em\lower.7ex\hbox{E}\kern-.125emX}}
\usepackage{makecell}
\usepackage{xurl}

\newtcolorbox{promptbox}[1]{
  breakable,
  colback=gray!5!white,
  colframe=gray!75!black,
  fonttitle=\bfseries,
  fontupper=\small,
  title=#1,
  sharp corners=southwest,
  boxrule=1pt,
  arc=4pt,
  left=6pt,
  right=6pt,
  top=6pt,
  bottom=6pt
}

\usepackage{xcolor}
\usepackage{soul,color}

\newcommand{\ourwork}{FlowGuard\xspace}

\begin{document}

\title{\ourwork: From Signals to Evidence for MCP Security Detection}

\author{
\IEEEauthorblockN{
Baichao An,
Pei Chen,
Geng Hong,
Yueyue Chen,
Mengying Wu
}

\IEEEauthorblockA{
Fudan University, Shanghai, China\\
\{bcan20, peichen19, ghong\}@fudan.edu.cn, \{yueyuechen25, wumy21\}@m.fudan.edu.cn
}
}


\maketitle

\begin{abstract}

The Model Context Protocol (MCP) enables LLM agents to interact with external tools through metadata exchange, tool invocation, and response consumption. 
Existing MCP security scanners primarily reason about suspicious semantic signals rather than real execution behaviors, which can lead to unreliable risk assessment.
For example, credential-like strings may simply be placeholders rather than actual leakage.
This gap requires runtime evidence for execution-related risks and careful semantic analysis for risks carried in metadata or returned content.

We present \ourwork, an evidence-grounded MCP security detection system. 
\ourwork combines semantic risk triage, recon-guided payload narrowing, schema-valid probe generation, evidence adjudication, and history-guided refinement. 
It verifies execution-related risks through runtime evidence and detects semantic risks in tool metadata and returned content.
We evaluate \ourwork on an executable benchmark containing 1,880 MCP cases across five vulnerability categories. 
\ourwork achieves F1 scores of 0.879 and 0.942 on the execution-related Command Injection and File System Access categories, respectively. 
Compared with existing dynamic scanners, \ourwork reduces end-to-end latency by up to 2.23$\times$. 
In the real-world evaluation, \ourwork reports 523 findings across 326 servers.
These results show that evidence-grounded detection can assess both execution-related and semantic risks in MCP interactions.

\end{abstract}


\section{Introduction}
\label{sec:introduction}

Large language models are rapidly evolving from passive text generators into autonomous agents that can invoke external tools, access proprietary data, and operate over real-world system resources. 
As the agent ecosystem expands, the Model Context Protocol (MCP) has emerged as a unified standard for connecting LLM agents with external tools and services, and has become a foundational component of modern AI agent ecosystems and application supply chains. 
By standardizing Session Initialization, Tool Discovery, Tool Invocation, and Response Consumption, MCP substantially lowers the barrier for integrating third-party capabilities and accelerates the deployment of tool-driven AI systems.

However, MCP also fundamentally expands the attack surface of LLM systems. Unlike traditional LLM applications that are largely confined to text input and output, MCP agents directly interact with file systems, databases, cloud services, DevOps platforms, and network interfaces through external tools. A flawed or malicious MCP server may therefore expose risky metadata, route parameters into unsafe backend operations, or return content that contaminates downstream agent reasoning.

Existing MCP scanners can already detect large numbers of suspicious signals from metadata, parameters, or runtime outputs, but these signals can be misleading. 
For example, credential-like strings may simply be placeholders rather than actual leakage, while command-like outputs may originate from user-input reflection or syntax errors instead of real command execution success. 
As a result, signal-based scanners often struggle to distinguish suspicious outputs from real security risks.
This limitation appears across existing MCP security approaches. 
Static analysis lacks runtime behavior observation (e.g., MCPScan~\cite{mcpscan-antgroup}); metadata auditing cannot determine runtime reachability (e.g., Agent-Scan~\cite{mcp_scan2025}); and dynamic methods often rely on broad payloads and weak response feedback, limiting their ability to confirm the vulnerability evidence (e.g., A.I.G~\cite{Tencent_AI-Infra-Guard_2025}). 
As a result, MCP security assessment must distinguish verified execution behavior from semantic content that may influence an agent.

This gap motivates an evidence-grounded scanner that combines semantic interpretation of tool metadata, schema-valid invocation, runtime feedback, and response attribution, using LLMs only for ambiguous semantic steps. Although related to black-box fuzzing and web security scanning, MCP requires separate treatment of backend execution evidence and content that may influence downstream LLM agents.

\textbf{Challenges.} 
Runtime MCP security detection therefore raises three core challenges: 
(1) Probes must simultaneously satisfy schema validity and attack effectiveness. A malformed probe is rejected before reaching backend logic, while a benign valid input may reveal little about exploitability. 
(2) Scanners must decide where to spend their probing budget under limited semantic visibility. Parameters such as \texttt{path}, \texttt{query}, or \texttt{url} look similar at the schema level even when they feed very different backend operations.
(3) Runtime responses are inherently ambiguous, since suspicious outputs may reflect user input, defensive behavior, or genuine evidence of exploitation.

\textbf{Key Insight.}
Our key insight is that MCP exposes complementary signals for two forms of security assessment. Tool Discovery metadata supports semantic risk detection and provides priors for runtime probing, while Tool Invocation and Response Consumption can provide concrete execution or disclosure evidence.
During Tool Invocation, runtime failures are not merely noise: schema-compliant probes may reveal backend fingerprints such as language, database engine, operating system, or path conventions. During Response Consumption, returned content provides candidate evidence, but that evidence must be adjudicated against source attribution and expected tool behavior.

\textbf{Our Work.}
Based on this insight, we design \ourwork, an evidence-grounded MCP security detection system. It verifies execution-related findings through runtime probing and identifies semantic risks in tool metadata and returned content without assuming that a downstream agent follows the content.
LLMs are used as bounded semantic subroutines in this process. The core design is the structured verification loop that constrains probing and decisions with MCP schemas, risk states, runtime feedback, and evidence-adjudication rules.

We evaluate \ourwork on an executable MCP security benchmark containing 1,880 cases across five vulnerability categories. 
\ourwork detects both execution-related and semantic risks, reaching F1 scores of 0.879 on Command Injection and 0.942 on File System Access while reducing mean latency by up to 2.23$\times$.
We further evaluate \ourwork on 8,000 real-world MCP servers from MCPZoo. 
\ourwork reports 523 findings across 326 servers.
These findings show the value of separating verified runtime evidence from semantic risk signals.

\textbf{Contributions.}
We make the following key contributions:

\begin{itemize}

    \item We design and implement \ourwork, an evidence-grounded MCP scanner that combines semantic risk detection with schema-valid probing, runtime feedback, and evidence adjudication.

    \item We construct an executable MCP security benchmark spanning five vulnerability categories, enabling systematic evaluation of black-box MCP security scanners under realistic runtime interaction settings.

    \item We evaluate \ourwork on the benchmark and 8,000 real-world MCP servers, covering detection effectiveness, runtime overhead, and report quality.
\end{itemize}

\section{Background}
\label{sec:background}

\subsection{MCP Overview}
\label{sec:mcp-protocol-overview}

The Model Context Protocol (MCP) defines a common interface for connecting applications, models, tools, and data sources~\cite{modelcontextprotocol}. Its message layer encodes communication in JSON-RPC 2.0~\cite{jsonrpc20}, while its transport layer supports stdio, SSE, and Streamable HTTP. 
An MCP deployment typically involves three roles: the host application, the MCP client, and one or more MCP servers. 
The host application, such as an AI assistant, editor, or IDE, manages the user-facing interaction and decides when external capabilities should be made available to the model. 
The MCP client implements the protocol on behalf of the host, while each MCP server exposes external capabilities through standardized primitives such as tools, resources, and prompts. 
Among these primitives, tools are the most security-critical for our study because they allow the agent to invoke server-side functionality with structured arguments and receive execution results.
A typical tool interaction proceeds through four conceptual stages: \emph{Session Initialization}, \emph{Tool Discovery}, \emph{Tool Invocation}, and \emph{Response Consumption}. We use these stages to describe the protocol workflow, while \ourwork focuses on the latter three stages, where tool metadata, invocation parameters, and returned content are exposed.

\begin{figure}[t]
    \centering
    \includegraphics[width=\linewidth]{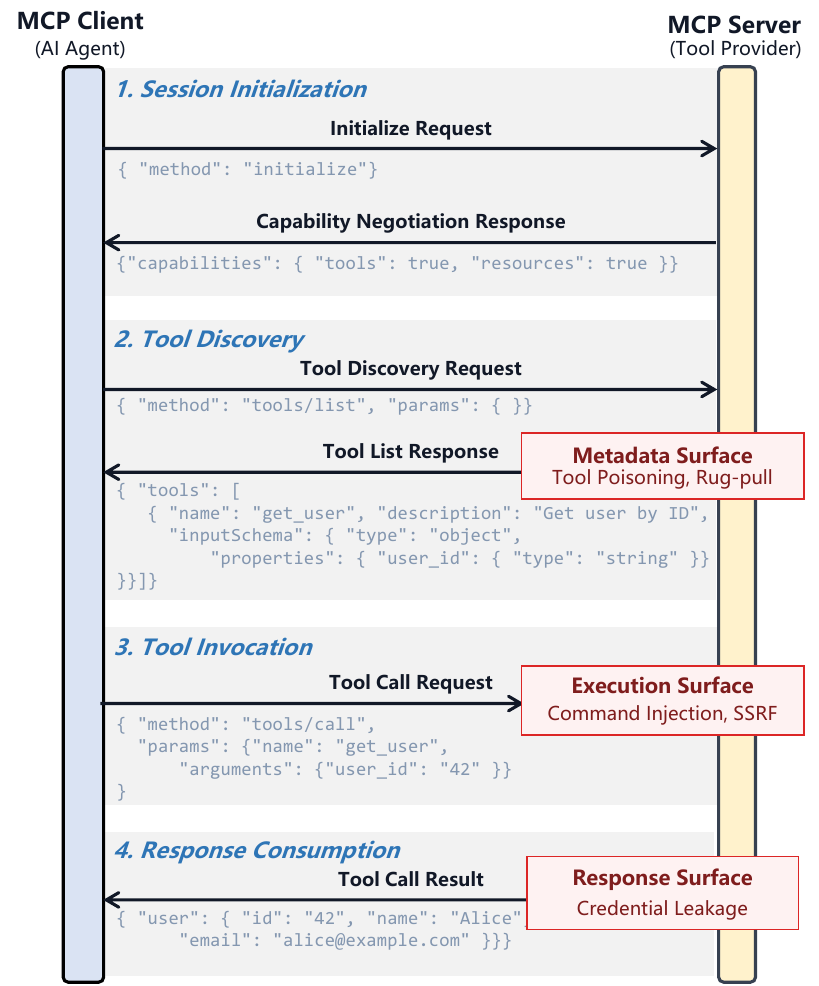}
    \caption{
    MCP interaction workflow and the attack surfaces analyzed by \ourwork during Tool Discovery, Tool Invocation, and Response Consumption. 
    }
    \label{fig:mcp_attack_surface}
\end{figure}

\subsection{Attack Surfaces in MCP Interaction}
\label{sec:security-threats-in-the-MCP-ecosystem}

As illustrated in Figure~\ref{fig:mcp_attack_surface}, Session Initialization establishes the protocol context, while \ourwork analyzes security risks exposed during Tool Discovery, Tool Invocation, and Response Consumption. These stages expose metadata, parameters, and returned content to the agent.

\noindent\textbf{Session Initialization.}
During this stage, the client establishes a session with the server and negotiates their supported capabilities.
We treat this as a setup stage rather than a detection target because it carries less attacker-controlled semantic content than later tool interactions.

\noindent\textbf{Tool Discovery.}
During this stage, the client requests the available tool list, through which it obtains each tool's name, natural-language description, and parameter schema.
These metadata are often incorporated into
the LLM context and used to guide tool selection and subsequent reasoning. 
An attacker can embed hidden instructions in a tool description to manipulate the LLM's behavior, commonly known as tool poisoning; register a malicious tool with a confusingly similar or overlapping interface to a legitimate one, known as tool shadowing; or change a tool's behavior after it has passed an initial review, known as a rug-pull attack. 
Existing defenses have made progress on metadata auditing through rule-based matching, static inspection, and LLM semantic analysis.

\noindent\textbf{Tool Invocation.}
In this stage, the client sends a structured invocation request.
Tool parameters may be propagated into backend sinks such as file systems, databases, command interpreters, or network clients, leading to path traversal, SQL injection, command execution, or SSRF. These risks often cannot be verified from descriptions and schemas alone, because they depend on how the server handles parameters at runtime. 
Dynamic analysis can exercise this stage through real invocations, but confirming exploitability requires probes that satisfy tool constraints and response analysis that can use runtime feedback.

\noindent\textbf{Response Consumption.}
In this stage, the client receives the corresponding execution result.
Tool call results are consumed by the LLM and may shape its subsequent actions. An attacker can embed malicious instructions in a tool response to hijack later agent behavior through indirect prompt injection, or cause the response to expose sensitive information such as system paths, database schemas, and API keys. 
Unlike metadata auditing, this level requires runtime interpretation: the detector must distinguish genuine leaks or injections from benign errors and normal tool outputs.

Overall, existing work has built substantial detection capability at the Tool Discovery stage, but important gaps remain at the Tool Invocation and Response Consumption stages. 
Metadata-stage risks can often be flagged before execution, but invocation-stage and response-stage risks require runtime evidence because their exploitability depends on backend data flow and returned content.

\subsection{Threat Model}
\label{sec:threat-model}

\subsubsection{System Architecture and Trust Boundaries}
\label{sec:system-architecture-trust-boundaries}
We consider a typical MCP deployment in which a host application (e.g., Claude Desktop~\cite{claude-desktop} or Cursor~\cite{cursor-ide}) embeds an MCP client that connects to one or more MCP servers. 
The host application, embedded MCP client, user, underlying LLM, and \ourwork scanning environment are trusted, while MCP servers are outside the trust boundary. 
This assumption reflects \ourwork's role as a server-side security scanner: it runs under the operator's authority and attributes evidence from standard MCP interactions to external servers. 
If the host, client, or scanner is compromised, probes, metadata, responses, or reports may be tampered with, invalidating server-side attribution. 
Such endpoint-compromise scenarios are outside our scope and require complementary defenses such as hardening, sandboxing, attestation, or independent traffic auditing.

\subsubsection{Attacker Model}
\label{sec:attacker-model}
We model two kinds of server behavior, rather than trying to decide whether the server operator is malicious.

\textit{Behavior A: Malicious or Policy-Violating MCP Behavior.} A server may expose harmful tool metadata, schemas, or returned content, such as tool poisoning, misleading descriptions, prompt-injection payloads, or credential-related disclosures. \ourwork treats these artifacts as untrusted observations and checks whether they provide concrete security evidence under standard MCP interactions.

\textit{Behavior B: Vulnerable-but-Benign MCP Implementation.} A server may be benign in intent but vulnerable in implementation. In this case, tool parameters may reach dangerous backend operations, such as command execution, path traversal, unauthorized file reads, SQL queries, or SSRF. \ourwork checks whether schema-valid invocations can produce observable evidence of such behavior.

In both cases, \ourwork verifies observable evidence rather than deciding why the server exposes the behavior.

\subsubsection{Attacker Capabilities}
\label{sec:attacker-capabilities}
Session Initialization provides connection and capability context, but is not itself a detection target. Our analysis treats server-provided metadata, execution behavior, and returned content as untrusted observations during Tool Discovery, Tool Invocation, and Response Consumption. During a scan, we assume the server does not try to recognize \ourwork, hide evidence only from the scanner, or change behavior only to evade probing. Servers that adapt to the scanner, delay malicious behavior across sessions, or behave differently for benign-looking clients are outside our current threat model.

\subsubsection{Scope and Limitations}
\label{sec:scope-limitations}
We focus on security risks exposed through standard MCP interactions, particularly high-risk behaviors triggered by tool parameters or returned outputs.
Our analysis targets MCP servers accessible through standard runtime interfaces and evaluates whether suspicious interaction signals correspond to real security-relevant behaviors under runtime execution.
The four stages describe the protocol workflow, but \ourwork detects risks only in Tool Discovery, Tool Invocation, and Response Consumption. The five evaluated categories are representative risks in these stages rather than a complete classification of MCP security.
We do not consider scenarios in which the host application, embedded MCP client, or \ourwork execution environment is already compromised.
We also exclude vulnerabilities that require white-box code analysis, local system privileges, adaptive probe evasion, long-term cross-session behavior, or attacks that bypass MCP tool interaction entirely, such as standalone social-engineering attacks unrelated to tool invocation.

\section{Methodology}
\label{sec:methodology}

\begin{figure*}[t]
    \centering
    \includegraphics[width=0.85\linewidth]{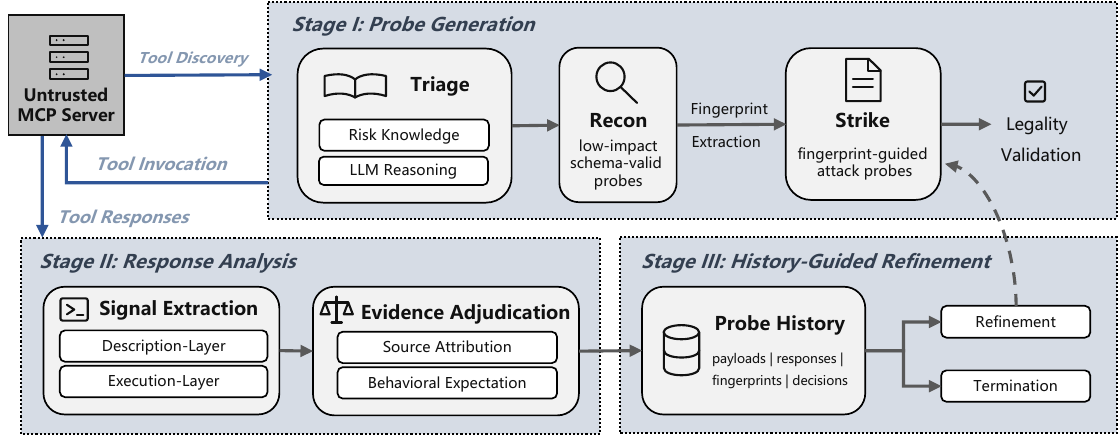}
    \caption{An overview of \ourwork.}
    \label{fig:overview}
\end{figure*}

\subsection{Challenges and Key Ideas}
\label{sec:challenges-key-ideas}

\textbf{\textit{Challenge 1: Probes must be both schema-valid and attack-effective, but these two goals conflict without backend visibility.}}
A useful security probe must satisfy two conflicting requirements. It must conform to the tool's type and format constraints, otherwise it will be rejected before reaching the backend. At the same time, it must carry a payload capable of triggering vulnerability behavior. Without knowledge of the backend implementation, constructing such probes is non-trivial. 
We decouple probing into two phases: a reconnaissance phase that extracts backend signals using low-impact, schema-compliant inputs, followed by a targeted phase that generates exploit-oriented probes guided by the inferred backend context.

\textbf{\textit{Challenge 2: Parameters with similar schemas can have different security semantics, yet probing should be efficient and effective.}}
At the schema level, parameters such as \texttt{path}, \texttt{query}, or \texttt{url} appear similar regardless of whether they are used in benign operations or passed directly into dangerous backend calls.
This semantic ambiguity makes it difficult to identify which inputs are security-relevant.
At the same time, exhaustively probing all parameters is inefficient, increases unnecessary load on MCP servers, and reduces the likelihood of triggering real vulnerabilities under a limited probing budget.
In the triage phase, we combine tool descriptions, parameter names, and JSON schemas to rank parameters by semantic risk. This reduces the probe space from all parameters to a small subset that is most likely to reach unsafe operations.

\textbf{\textit{Challenge 3: Tool responses do not provide a reliable signal for distinguishing benign failures from true vulnerabilities.}}
Responses by MCP tools may correspond to schema validation failures, normal business logic, or genuine vulnerability triggers, and these cases often appear similar. Error formats vary widely across backend environments, making simple pattern matching unreliable, while overly conservative interpretation leads to missed findings. 
Relying on single-response heuristics can therefore lead to both false positives and missed vulnerabilities.
We maintain a local history of probes and responses, 
and iteratively refine both probing strategies and vulnerability hypotheses, enabling the system to accumulate evidence over multiple rounds rather than relying on any single response.

\subsection{System Overview}
\label{sec:system-overview}
Our system operates as a closed-loop active probing pipeline, as illustrated in Figure~\ref{fig:overview}. \ourwork performs semantic risk detection for metadata and returned instructions, and runtime evidence verification for execution-related risks. It takes Tool Discovery metadata from an MCP server as input and passes it through five phases: \textsc{Triage}, \textsc{Recon}, \textsc{Strike}, \textsc{Analysis}, and \textsc{Refinement}. The pipeline iterates over these structured states until it reaches a verdict, exhausts the probe budget, or observes no actionable new signal. Appendix~\ref{app:prompt-templates} lists the prompts used to instantiate the bounded semantic subroutines.

\textsc{Triage} analyzes tool names, descriptions, and parameter schemas from Tool Discovery, and outputs candidate tool, parameter, and risk tuples for probing.

\textsc{Recon} sends low-impact, schema-compliant probes to selected parameters and extracts environment fingerprints such as backend language, database engine, and path conventions.

\textsc{Strike} uses these fingerprints and schema constraints to issue targeted Tool Invocation probes that are more likely to reach security-relevant backend paths.

\textsc{Analysis} extracts candidate response signals and applies source-attribution and behavioral-expectation checks to produce candidate signals or confirmed evidence.

\textsc{Refinement} updates the probe history and decides whether to continue probing, stop as inconclusive, or emit a final report.
The following subsections detail how these states are constructed, validated, and refined during the scan.

\subsection{Probe Generation}
\label{sec:probe-generation}

This subsection describes target selection and schema-valid probe construction.

\subsubsection{Recon-Guided Payload Narrowing}

Effective attack payloads are highly environment-dependent. For example, command-injection syntax for a Python backend differs from that for Node.js, and schema-enumeration queries for SQLite differ from those for MySQL. In a black-box setting where the backend is unknown, issuing strike probes directly expands the payload space into broad enumeration, resulting in low hit rates. Prior black-box fuzzing work has shown that implementation clues exposed in target responses are critical for specializing subsequent inputs~\cite{feng2021snipuzz}. We therefore split probing into a \textsc{Recon} phase and a \textsc{Strike} phase. A low-impact probe is first sent to extract environment signals from a single call, and the resulting fingerprints are then used to select a small number of high-yield payloads, narrowing the candidate payload space from broad enumeration to a targeted set.

\textbf{Recon Probe Design.}
For each threat category, a low-impact fault value triggers a controlled, information-rich failure. 
For example, file-system parameters may use non-existent paths to elicit errors, and command-execution parameters may include minimal shell markers to test interpretation.
The goal of a \textsc{Recon} probe is not to complete a successful Tool Invocation, but to obtain responses that reveal path formats, language frameworks, database engines, or system error messages at minimal cost.

\textbf{Fingerprint Extraction.}
After receiving a \textsc{Recon} response, we apply rule-based extraction to identify environment fingerprints, including the operating system, backend language, and database type. For example, \texttt{traceback} or \texttt{django} suggests a Python backend; \texttt{goroutine} or \texttt{panic} suggests Go; \texttt{sqlite3} suggests SQLite; and \texttt{C:\textbackslash} or \texttt{win32} suggests Windows. If rule-based extraction is incomplete, a bounded semantic fingerprinter returns the missing fields as a structured JSON object. The extracted fingerprints are then used to select and rank payloads for the \textsc{Strike} phase.

\textbf{Strike Probe Generation.}
The extracted fingerprints are used to generate a single threat-specific strike payload for each high-risk parameter and threat family. 
For example, command-execution parameters leverage backend language and OS to craft appropriate shell commands, file-like parameters preserve a benign prefix before adding minimal shell markers, and database parameters may query platform-specific tables. 
For cases not covered by deterministic templates, the generator uses the structured fingerprint, schema constraints, and threat family to produce one bounded candidate payload. The payload is coerced to the declared parameter type before being embedded into an otherwise safe base invocation.
All strike payloads are additionally constrained to avoid state-modifying side effects, external network contact, and destructive operations such as writes or deletes.

\textbf{Probe Legality.}
Tool Invocation requests are constrained by JSON Schema fields such as
\texttt{type}, \texttt{enum}, \texttt{pattern}, and \texttt{maxLength}. 
Non-conforming probes may be rejected by the MCP schema-validation layer before
they reach the vulnerable backend path. FlowGuard therefore makes probe generation
schema-aware. During triage, it extracts parameter types and explicit constraints
from the tool schema and includes them in the structured context used for payload
generation. During probing, candidate payloads are coerced or wrapped according
to the declared parameter type: numeric and boolean parameters require parseable
values, arrays receive schema-shaped elements, object parameters require structured
objects, and string parameters preserve the attack intent when possible. If a
probe is rejected by schema validation, the validation response is fed back into
the refinement loop to generate a more compatible follow-up probe.

\subsubsection{Semantic Risk Triage}

Before Tool Invocation, an MCP server already exposes tool descriptions, JSON schemas, and parameter names during Tool Discovery. These static signals provide useful priors about backend behavior. For example, parameters such as \texttt{cmd}, \texttt{command}, and \texttt{exec} often indicate command execution risk; \texttt{file}, \texttt{path}, and \texttt{dir} suggest file system access; and \texttt{url}, \texttt{host}, and \texttt{callback} commonly appear in network-request contexts~\cite{chow2023beware}. 
The goal of \textsc{Triage} is to use these signals to focus probing on a small set of high-risk parameters, reducing exploration cost while retaining potentially impactful targets.

\textbf{Risk Knowledge Construction.}
The risk knowledge base provides fast-path rules for three parameter patterns: command-like inputs, file-like inputs, and prompt-bearing inputs. It accelerates routing for clear parameter semantics rather than defining all risks supported by \ourwork.
Each entry is assigned a risk score based on the strength and consistency of evidence from CVE-derived parameter names, Fluffy's parameter to sink observations, and the OWASP Top 25 Parameters list.
Parameters with strong and unambiguous scores are selected as fast path parameters and can be routed directly to the corresponding probe type.
Table~\ref{tab:risk_knowledge_base} summarizes the resulting knowledge base.
Prompt-bearing inputs are an internal routing pattern and are distinct from the response-layer Prompt Injection category evaluated in our benchmark.

\textbf{Risk Vectors and Fast-Path Routing.}
Each parameter is represented as a risk vector over the three fast-path patterns in the knowledge base.
As summarized in Table~\ref{tab:risk_knowledge_base}, parameters with strong and unambiguous signals are routed to a fast path, where they directly receive probes for the corresponding risk type.
Parameters with weaker or conflicting signals enter an uncertainty path rather than being immediately pruned. In this path, \ourwork retains parameters whose names, descriptions, schema constraints, or cross-parameter composition suggest possible high-impact behavior, while pruning only parameters with consistently low-risk evidence.
Risks not covered by the fast path, including Credential Leakage, are retained through uncertainty-path triage or identified from returned content during Response Analysis.

\textbf{Uncertainty-Path Triage.}
For parameters that the fast path cannot resolve, \ourwork invokes a bounded semantic triage step with structured JSON output. This step infers tool intent, parameter roles, cross-parameter composition, and risk labels from concrete metadata evidence; its output is used only to choose among high-confidence routing, uncertainty retention, and low-risk pruning. Appendix~\ref{app:prompt-templates} provides the implementation prompt and expected output fields.

\subsection{Response Analysis}
\label{sec:response-analysis}

\subsubsection{Signal Extraction from MCP Responses}
Security signals in an MCP workflow are not limited to tool execution results. Before Tool Invocation, the agent obtains tool names, descriptions, and JSON schemas during Tool Discovery, and this metadata may directly affect tool selection and argument construction. We therefore separate signal identification into two categories: metadata-layer signals from Tool Discovery, and execution-layer signals from Tool Invocation and Response Consumption.

\textbf{Description-Layer Detection.}
Tool Discovery metadata is itself an attack surface.
A malicious MCP server can embed hidden instructions in tool descriptions or schemas, influencing the agent before Tool Invocation. To balance efficiency and coverage, we use a three-tier pipeline. First, name-confusion detection normalizes tool names by lowercasing, removing non-alphanumeric characters, and applying common leet substitutions, then compares them against high-risk authoritative tool names. Second, fast rule matching applies regular expressions to the concatenated text of tool names, descriptions, and key schema fields to detect hidden instructions, cross-tool control, schema manipulation, and Unicode obfuscation. Third, when explicit rules do not fire, a bounded semantic check returns structured labels for subtler poisoning semantics. Only unresolved cases reach this step, and its labels are passed to evidence adjudication rather than emitted as findings.

\textbf{Execution-Layer Detection.}
During Tool Invocation and Response Consumption, different vulnerability types leave different observable signals. For credential leakage, we use rule libraries such as Gitleaks to match high-confidence structured credential patterns, including API keys, tokens, private keys, cloud credentials, and configuration fragments. For command injection, we extract explicit execution evidence from responses, such as \texttt{id} output, system file fragments, OS version strings, shell paths, and command-not-found errors. These patterns are not used as standalone proof of a vulnerability; they only mark response fragments that may indicate interaction with a dangerous backend context.

Signal identification is therefore a high-recall evidence collection step rather than a final decision procedure. Rule-matched fragments are retained as inputs to the subsequent \textsc{Evidence Adjudication} step. Conversely, responses without rule matches are not immediately considered safe, since subtle leaks, blind execution, or context-dependent failures may only become meaningful when analyzed together with the probe, tool metadata, and probe history. \ourwork therefore treats suspicious fragments as candidate signals rather than findings. A candidate becomes vulnerability evidence only after it is attributed to backend behavior and judged inconsistent with the tool's expected semantics.
Thus, signal extraction can raise suspicion, but it never emits a finding by itself.

\subsubsection{Evidence Adjudication}

In a black-box MCP setting, the scanner can only observe tool outputs without access to backend implementations or execution paths. As a result, the presence of a suspicious signal does not necessarily indicate a real vulnerability. Determining whether an observed response constitutes a true security finding requires reasoning beyond the response itself.

We perform adjudication along two dimensions of evidence.

\textbf{E1: Source Attribution.}
This dimension determines whether suspicious content in the response is produced by the backend or merely reflects the probe input. If the content is an echo of attacker-controlled input, it does not indicate information disclosure. Only content independently generated by the system constitutes valid vulnerability evidence.
For example, a response that simply echoes a probe marker is not treated as credential leakage or command output, even if the marker matches a suspicious pattern.

\textbf{E2: Behavioral Expectation.}
This dimension determines whether the response can be explained by the tool's intended functionality or defensive mechanisms. Responses such as authentication failures, invalid input rejections, permission errors, and resource-not-found messages are consistent with expected behavior. Although they may appear anomalous, they reflect correct handling of malicious input rather than successful exploitation.
For example, a file-reading tool returning an allowed fixture file is expected behavior, whereas returning content outside its declared workspace boundary becomes evidence for a File System Access finding.

\textbf{Decision Outcome.}
For execution-related risks, the system classifies a signal as input reflection, normal defense, or confirmed runtime evidence. Reflections and normal defenses are fed back as constraints, while only backend-originated behavior outside the tool's expected boundary is confirmed. For Tool Poisoning and Prompt Injection, the output is instead a semantic-risk verdict; it does not claim that a downstream agent will follow the detected instruction. The adjudicator records the source, expected behavior, and verdict in structured fields. Appendix~\ref{app:prompt-templates} gives one implementation prompt for producing these fields.

\subsection{History-Guided Refinement}

An initial \textsc{Strike} probe does not always trigger a vulnerability directly, but even a failed response can provide useful information. Instead of treating a failed probe as merely invalid, we treat the response as feedback that constrains the next probe.

For each parameter and risk type, the system maintains a local probe history:
\[
H_t = \{(\mathbf{p}_1, r_1),(\mathbf{p}_2, r_2),\ldots,(\mathbf{p}_t, r_t)\},
\]
where \(\mathbf{p}_i\) is the payload sent in round \(i\), and \(r_i\) records the response, validation outcome, extracted signals, and adjudication result. The refinement module summarizes this history into constraints and either proposes the next payload or stops when no actionable signal remains.

\textbf{Distinguishing Failure Types.}
The refinement module distinguishes between two broad classes of failures. If the response indicates that the payload was rejected by schema or format validation, the next probe repairs the structural issue while preserving the original attack intent. If the payload passes schema validation but triggers a business-logic error, the system treats this as evidence that the input has reached part of the backend execution path and generates a more targeted follow-up. For example, a file-not-found error suggests that path parsing has occurred, so the next probe can adjust directory depth or target file names; a SQL syntax error suggests that the input has entered query construction logic, so the next probe can adjust quote escaping or comment characters.
The refinement step also consumes the adjudication outcome from the previous round: input reflections lead to payload-marker changes, normal defensive responses reduce the priority of that direction, partial backend clues trigger more specialized probes, and confirmed evidence terminates the loop with a report.

\textbf{Termination.}
The refinement loop stops when \ourwork obtains confirmed evidence, exhausts the probing budget \(B\), repeats an equivalent probe, or when the refinement module determines from the probe history that subsequent rounds provide no actionable new signal.
At that point, the refinement module emits a stop decision, and \ourwork terminates refinement for the current parameter and risk type. This loop lets later probes adapt to explicit schema constraints and runtime backend constraints observed from prior responses.
Algorithm~\ref{alg:flowguard-loop} gives the pseudocode for the loop.

\section{Benchmark Construction}
\label{sec:benchmark-construction}

\subsection{Benchmark Overview}

Traditional software-security benchmarks provide a useful model for constructing executable test suites with explicit ground truth. OWASP Benchmark~\cite{owaspbenchmark} and NIST SARD/Juliet~\cite{juliet}, for example, pair vulnerable and benign cases for evaluating program-analysis tools. However, they do not capture risks in MCP tool metadata, invocation parameters, and returned content, and therefore cannot directly evaluate scanner performance across Tool Discovery, Tool Invocation, and Response Consumption.

Several existing MCP benchmarks place an LLM agent in an attack scenario and measure whether the attack induces unsafe agent behavior. MSB~\cite{zhang2026mcpsecuritybenchmsb}, MCPSecBench~\cite{yang2026mcpsecbenchsystematicsecuritybenchmark}, and MCP-SafetyBench~\cite{zong2026mcpsafetybenchbenchmarksafetyevaluation} follow this evaluation setting. Their outcomes therefore depend on the agent model, prompt, and execution policy in addition to the MCP server.
\ourwork evaluates a different target: the security scanner. Each benchmark case must therefore provide a known positive or negative label together with explicit evidence for determining whether the scanner's report is correct. This ground truth enables scanner-level precision, recall, and false-positive measurement, but is not systematically provided by agent-behavior benchmarks. The two evaluation settings are complementary: one measures the downstream effect of an attack on an agent, while the other measures whether a scanner can identify the corresponding risk from MCP metadata, invocations, and responses.

\begin{table*}[t]
\centering
\caption{Overview of the MCP security benchmark. Threat Stage indicates where in the MCP interaction flow the attack is introduced.}
\label{tab:benchmark}
\resizebox{\textwidth}{!}{
\setlength{\tabcolsep}{4pt}
\begin{tabular}{llcrrc}
\toprule
\textbf{Category} & \textbf{Threat Stage} & \textbf{Representative Patterns} & \textbf{Total} & \textbf{Pos.} & \textbf{Neg.} \\
\midrule
\textbf{Command Injection} & Tool Invocation
& shell construction, eval, config dispatch, staged execution 
& \textbf{360} & 144 & 216 \\

\textbf{Tool Poisoning} & Tool Discovery
& explicit/implicit hijacking, parameter tampering, benign metadata 
& \textbf{550} & 485 & 65 \\

\textbf{Prompt Injection} & Response Consumption
& unsolicited instruction, fake error, indirect injection 
& \textbf{500} & 300 & 200 \\

\textbf{Credential Leakage} & Response Consumption
& filesystem, runtime, database, cloud, SCM leakage 
& \textbf{220} & 180 & 40 \\

\textbf{File System Access} & Tool Invocation
& traversal, absolute path, symlink, archive extraction 
& \textbf{250} & 200 & 50 \\

\midrule
\multicolumn{3}{l}{\textbf{Total}} 
& \textbf{1,880} & \textbf{1,309} & \textbf{571} \\
\bottomrule
\end{tabular}
}
\end{table*}

To address these gaps, we construct an executable MCP security benchmark spanning five categories: \emph{Command Injection}, \emph{Tool Poisoning}, \emph{Prompt Injection}, \emph{Credential Leakage}, and \emph{File System Access}. 
These categories cover common high-risk behaviors observed in real MCP ecosystems across Tool Discovery, Tool Invocation, and Response Consumption, including tool metadata exposure, unsafe parameter execution, and untrusted runtime outputs.
The benchmark contains two task groups. Command Injection, Credential Leakage, and File System Access are execution-evidence risks labeled by observable execution or disclosure. Tool Poisoning and Prompt Injection are semantic-content risks labeled by adversarial intent in metadata or returned content; they do not establish that a downstream agent will follow the instruction.
The benchmark construction follows established security specifications adapted to MCP, including CWE-78~\cite{cwe78}, CWE-22~\cite{cwe22}, OWASP WSTG~\cite{owaspwstg}, MCPTox-derived poisoning patterns, and the positive-negative pairing style of OWASP Benchmark and SARD/Juliet~\cite{owaspbenchmark,juliet}. Each sample consists of a runnable MCP server, defined attack triggers, ground-truth labels, and verification evidence, enabling quantitative evaluation of both detection accuracy and false-positive rates.

Table~\ref{tab:benchmark} summarizes the benchmark composition. In total, the benchmark contains 1,880 executable MCP cases, including 1,309 positive cases and 571 negative or hard-negative cases. The distribution is intentionally not uniform across categories: semantic attacks such as tool poisoning and prompt injection contain more variants due to their diverse textual realizations, while execution-path vulnerabilities are organized around representative backend patterns and paired safe implementations.
Ground-truth labels are assigned by benchmark design rather than by scanner output. A positive case has a defined trigger and category-specific evidence, such as command markers, fixture file contents, synthetic canaries, or instruction patterns in metadata or responses. A negative or hard-negative case preserves similar surface semantics, high-risk names, or suspicious-looking outputs, but its safe implementation should not produce the corresponding evidence.
We further validate the benchmark with category-specific checks.
For Command Injection, File System Access, and Credential Leakage, we run each generated MCP server with both benign and trigger inputs.
These checks confirm that benign inputs do not expose evidence, positive triggers produce the expected marker or synthetic canary, and paired safe variants block the same evidence.
For tool poisoning and prompt injection, we check the generated metadata and templates: positive cases contain the intended instruction pattern, while benign and hard-negative cases keep similar surface language but omit that pattern.

\subsection{Category-Specific Design}

We construct each category according to where the risk appears in the three stages analyzed by \ourwork. Execution-path vulnerabilities model unsafe flows from Tool Invocation parameters to backend operations, while semantic risks appear in Tool Discovery metadata or Response Consumption content. Each category includes positive cases and negative or hard-negative cases with similar surface semantics.

\textbf{Command Injection.}
Command injection occurs when user-controlled input is interpreted as part of an operating-system command, allowing an attacker to alter the intended command semantics or execute unintended commands. Following CWE-78~\cite{cwe78}, we generate MCP tools whose parameters flow into command-execution patterns. Positive cases cover direct shell construction, string concatenation, f-string command construction, configuration-driven shell dispatch, and environment-variable expansion. Negative cases preserve similar business semantics but use safer argument-vector execution, validated argument-vector calls, fixed or constant command templates, quoted shell arguments, timeout-bounded execution, or documentation-only command examples, following the positive-negative pairing style of OWASP Benchmark and SARD/Juliet.

\textbf{Tool Poisoning.}
Tool poisoning refers to attacks in which malicious instructions are embedded in tool metadata, such as tool names, descriptions, or parameter schemas, so that an LLM agent is induced to misuse a tool or change its intended behavior before Tool Invocation. A positive case must contain metadata that attempts to alter tool selection, parameter construction, privilege assumptions, or cross-tool behavior; benign controls may contain rich tool descriptions without such steering intent. This category targets the Tool Discovery stage, where the agent first observes tool metadata and incorporates it into planning. We adapt tool poisoning cases from MCPTox~\cite{wang2025mcptoxbenchmarktoolpoisoning}, convert the original poisoned tool descriptions into executable FastMCP servers, and preserve their template/paradigm annotations and risk categories. We additionally generate P0 benign tool-metadata controls to measure scanner false positives, which is different from ASR-oriented agent robustness evaluation.

\textbf{Prompt Injection.}
Prompt injection refers to attacks in which untrusted textual content is crafted to override, manipulate, or conflict with the agent's original instructions. In our benchmark, this category focuses on indirect prompt injection through tool outputs rather than direct user prompts. The malicious content is returned during Response Consumption and may be mistaken by the agent as trustworthy execution context. Positive cases inject exfiltration-oriented instructions into normal outputs, forged error messages, or simulated external data sources, corresponding to unsolicited-instruction, false-error, and remote-injection paradigms. Negative cases include clean outputs and hard negatives that contain instruction-like text, logs, code fragments, or command snippets but do not ask the agent to change task goals or violate its authorization boundary.

\textbf{Credential Leakage.}
Credential leakage refers to the unintended exposure of secrets, authentication tokens, API keys, session identifiers, or other sensitive configuration values through tool execution results. In the MCP setting, such leakage can occur when a tool reads from sensitive runtime state, local files, databases, cloud or Kubernetes-style configuration, runtime command outputs, provider debug outputs, or application metadata and returns the content to the agent. Our samples use synthetic canary secrets as ground truth to avoid exposing real credentials. The canaries cover structured secret formats and configuration-like fragments, and a positive label requires canary exposure outside the tool's intended semantics. Negative cases return redacted values, look-alike tokens, safe placeholders, or explicitly sanitized outputs.

\textbf{File System Access.}
File system access vulnerabilities occur when user-controlled path input allows a tool to read, include, or otherwise operate on files outside the intended directory boundary. Following CWE-22 and OWASP WSTG~\cite{cwe22,owaspwstg}, we construct path-based MCP tools over local fixtures. The evaluated dataset focuses on read-evidence cases, covering traversal read, absolute-path read, symlink-follow escape, and template-include escape. Negative cases implement boundary checks, path normalization, symlink rejection, allowlisted roots, or template/file-name validation. The generator also supports side-effect families such as unsafe writes, deletes, and archive extraction, but these are not included in the default evaluated benchmark.

\section{Evaluation}
\label{sec:evaluation}

We evaluate \ourwork through the following five research questions:

\begin{itemize}
    \item \textbf{RQ1 (Detection Effectiveness):}  
How accurate is \ourwork in detecting different types of MCP vulnerabilities, and can it maintain high coverage while effectively controlling false positives?
    \item\textbf{RQ2 (Probe Quality and Efficiency):}  
Can the constraint-aware probe generation strategy produce schema-compliant inputs while improving probe success rate and overall detection efficiency?

    \item\textbf{RQ3 (Runtime Overhead):}  
What is the runtime overhead of \ourwork, and does its time and interaction cost meet the requirements of practical deployment?

    \item\textbf{RQ4 (Mechanism Contribution):}  
What are the contributions of key components, including semantic risk triage, the Recon and Strike probing strategy, and response analysis with attribution, to the overall system performance?

    \item\textbf{RQ5 (Real-World Effectiveness):}  
When applied to real-world MCP servers, what risks does \ourwork report, and how many sampled reports contain concrete runtime evidence?
\end{itemize}

\begin{table*}[t]
\centering
\caption{Detection effectiveness across five MCP vulnerability categories. Error denotes failed runs and is excluded from denominators.}
\label{tab:overall-detection}
\begin{tabular}{llrrrrrrrrr}
\toprule
\textbf{Category} & \textbf{Method} & \textbf{TP} & \textbf{FP} & \textbf{TN} & \textbf{FN} & \textbf{Error} & \textbf{Precision} & \textbf{Recall} & \textbf{F1} & \textbf{Accuracy} \\
\midrule

\multirow{4}{*}{Credential Leakage}
& \ourwork & 140 & 4 & 36 & 40 & 0 & 0.9722 & 0.7778 & 0.8642 & 0.8000 \\
& MCPScan & 0 & 0 & 40 & 180 & 0 & 0.0000 & 0.0000 & 0.0000 & 0.1818 \\
& MCP-Scanner & 13 & 1 & 39 & 167 & 0 & 0.9286 & 0.0722 & 0.1340 & 0.2364 \\
& A.I.G & 115 & 14 & 26 & 65 & 0 & 0.8915 & 0.6389 & 0.7443 & 0.6409 \\

\midrule
\multirow{4}{*}{Tool Poisoning}
& \ourwork & 480 & 0 & 65 & 5 & 0 & 1.0000 & 0.9897 & 0.9948 & 0.9909 \\
& MCPScan & 324 & 13 & 52 & 161 & 0 & 0.9614 & 0.6680 & 0.7883 & 0.6836 \\
& MCP-Scanner & 329 & 8 & 57 & 156 & 0 & 0.9763 & 0.6784 & 0.8005 & 0.7018 \\
& A.I.G & 327 & 49 & 16 & 158 & 0 & 0.8697 & 0.6742 & 0.7596 & 0.6236 \\

\midrule
\multirow{4}{*}{Command Injection}
& \ourwork & 113 & 0 & 216 & 31 & 0 & 1.0000 & 0.7847 & 0.8794 & 0.9139 \\
& MCPScan & 144 & 180 & 36 & 0 & 0 & 0.4444 & 1.0000 & 0.6154 & 0.5000 \\
& MCP-Scanner & 0 & 0 & 216 & 144 & 0 & 0.0000 & 0.0000 & 0.0000 & 0.6000 \\
& A.I.G & 69 & 41 & 165 & 66 & 19 & 0.6273 & 0.5111 & 0.5633 & 0.6862 \\

\midrule
\multirow{4}{*}{Prompt Injection}
& \ourwork & 281 & 6 & 194 & 19 & 0 & 0.9791 & 0.9367 & 0.9574 & 0.9500 \\
& MCPScan & 100 & 66 & 134 & 200 & 0 & 0.6024 & 0.3333 & 0.4292 & 0.4680 \\
& MCP-Scanner & 0 & 0 & 200 & 300 & 0 & 0.0000 & 0.0000 & 0.0000 & 0.4000 \\
& A.I.G & 252 & 72 & 104 & 30 & 42 & 0.7778 & 0.8936 & 0.8317 & 0.7773 \\

\midrule
\multirow{4}{*}{File System Access}
& \ourwork & 178 & 0 & 50 & 22 & 0 & 1.0000 & 0.8900 & 0.9418 & 0.9120 \\
& MCPScan & 190 & 48 & 2 & 10 & 0 & 0.7983 & 0.9500 & 0.8676 & 0.7680 \\
& MCP-Scanner & 2 & 1 & 49 & 198 & 0 & 0.6667 & 0.0100 & 0.0197 & 0.2040 \\
& A.I.G & 64 & 0 & 50 & 134 & 2 & 1.0000 & 0.3232 & 0.4885 & 0.4597 \\

\bottomrule
\end{tabular}
\end{table*}

\subsection{Experiment Setup}

\textbf{Evaluation Targets.}  
We deploy \ourwork as an active security scanner that interacts with target servers strictly through the standard MCP protocol.
For benchmark evaluation, each case is executed as an independent MCP server instance with its ground-truth trigger and label hidden from the scanner. 
RQ1 reports end-to-end detection results over all five benchmark categories. 
Subsequent mechanism-oriented analyses focus on Command Injection, Credential Leakage, and File System Access, because they directly exercise \ourwork's active parameter-level probing, refinement, runtime overhead, and evidence collection mechanisms, whereas Tool Poisoning and Prompt Injection are primarily metadata- or response-semantic threats.
For real-world evaluation, we interact with MCPZoo servers through MCPZoo's provided access interface.

\textbf{Baselines.}  
We compare \ourwork with three representative MCP security scanners covering different analysis mechanisms. 
For static source-code analysis, we use MCPScan, which analyzes MCP server implementations to identify potentially risky behaviors. 
For metadata-based analysis, we use MCP-Scanner, which inspects tool descriptions and related metadata for suspicious patterns. 
For dynamic interaction analysis, we use A.I.G (AI Infra Guard), which performs runtime probing by generating tool invocations and analyzing execution responses. 
These baselines collectively represent the major existing MCP scanning approaches and allow us to compare \ourwork across different detection mechanisms.

\textbf{Implementation and Configuration.}
\ourwork uses Qwen3-235B-A22B-Instruct as the default LLM across all experiments. 
Unless otherwise specified, the timeout for each scanning run is set to 900 seconds, and the maximum probing budget B is limited to 5 rounds. 
The probing process terminates earlier once sufficient evidence is collected or no additional useful signal is observed.
For fair comparison, all baseline scanners use their default or recommended configurations under the same LLM and overall timeout constraint.

\textbf{Metrics.}  
We evaluate performance using multiple metrics, including detection accuracy (Precision, Recall, and F1 score), false positive rate, probe validity, average number of probing rounds, and overall runtime cost. For real-world MCP servers, we additionally perform manual verification of selected findings to assess their practical validity.

\begin{table*}[t]
\centering
\caption{Probe quality comparison. Reached Calls reports the number of probes that reach backend execution paths, with the corresponding percentage in parentheses. Evidence Calls reports the number of reached probes that produce security-relevant evidence, with the corresponding percentage in parentheses.}
\label{tab:probe-quality}
\begin{tabular}{llrrrrrr}
\toprule
\textbf{Task} 
& \textbf{Method} 
& \textbf{Total Probes}
& \textbf{Reached Calls (\%)}
& \textbf{Evidence Calls (\%)}
& \textbf{TP} 
& \textbf{Probes/TP} \\
\midrule

\multirow{2}{*}{Command Injection}
& \ourwork & 2,048 & 2,044 (99.80\%) & 1,171 (57.29\%) & 113 & 18.12 \\
& A.I.G & 3,709 & 3,578 (96.47\%) & 660 (18.45\%) & 69 & 53.78 \\

\midrule
\multirow{2}{*}{Credential Leakage}
& \ourwork & 1,386 & 1,355 (97.76\%) & 407 (30.04\%) & 140 & 9.90 \\
& A.I.G & 1,857 & 1,800 (96.93\%) & 730 (40.56\%) & 115 & 16.14 \\

\midrule
\multirow{2}{*}{File System Access}
& \ourwork & 1,433 & 1,419 (99.02\%) & 1,199 (84.50\%) & 178 & 8.05 \\
& A.I.G & 2,721 & 2,642 (97.10\%) & 2,381 (90.12\%) & 64 & 42.52 \\
\bottomrule
\end{tabular}
\end{table*}

\subsection{RQ1: Detection Effectiveness}

This section evaluates whether \ourwork can verify execution-evidence risks and detect semantic-content risks while controlling false positives.
Table~\ref{tab:overall-detection} reports the results across five categories.
We compare \ourwork with A.I.G as the main runtime baseline, MCP-Scanner as a metadata-level baseline, and MCPScan as a complementary static baseline.

\ourwork achieves the best overall results among all baselines.
Compared with A.I.G, \ourwork achieves higher F1 scores in all five categories.
The largest improvements appear in Command Injection and File System Access, where \ourwork reaches F1 scores of 0.879 and 0.942, compared to 0.563 and 0.489 for A.I.G, respectively.
These gains are most pronounced for vulnerabilities that require runtime validation, where generic probing strategies often fail to confirm exploitability.
MCP-Scanner performs poorly on these execution-layer risks, obtaining zero recall on Command Injection and only 0.010 recall on File System Access, which shows the limitation of relying only on tool descriptions.
For the semantic-content categories, the results measure whether scanners recognize adversarial content rather than whether a downstream agent follows it.
For Tool Poisoning, \ourwork reaches near-perfect performance with an F1 score of 0.995, outperforming A.I.G and MCP-Scanner.
For Prompt Injection, \ourwork achieves the highest F1 score of 0.957, outperforming both A.I.G and MCP-Scanner.
For Credential Leakage, \ourwork improves recall from 0.639 to 0.778 and precision from 0.892 to 0.972 over A.I.G, demonstrating better discrimination between true secret exposure and benign or reflected content.
MCP-Scanner achieves high precision but very low recall on Credential Leakage, indicating that metadata-level matching can identify a few obvious cases but misses most runtime leakage behaviors.
Overall, \ourwork performs consistently across the two task groups while avoiding the broad false positives observed in static analysis.

\subsection{RQ2: Probe Quality and Efficiency}

This section evaluates whether \ourwork generates high-quality probes rather than merely issuing more probes.
We report five metrics: 
\textit{Total Probes}, the number of probes sent; 
\textit{Reached Calls}, the number of probes that reached backend execution paths; 
\textit{Evidence Calls}, the number of reachable probes that induced security-relevant signals; 
\textit{True Positives (TP)}, the number of vulnerabilities triggered; 
and \textit{Probes/TP}, the mean number of probes required per true positive.
The Reached Calls and Evidence Calls metrics include both absolute counts and percentages.

As shown in Table~\ref{tab:probe-quality}, \ourwork achieves substantially higher probe efficiency across the three execution-evidence categories. 
For example, on Command Injection, it generates 113 true positives using only 2,048 probes (18.12 probes per TP), compared with A.I.G, which produces 69 true positives with 3,709 probes (53.78 probes per TP). 
Similarly, for Credential Leakage and File System Access, \ourwork requires far fewer probes per detected vulnerability while still inducing a high fraction of evidence, demonstrating that effective probe design and targeting are more important than total probe volume.

\subsection{RQ3: Runtime Overhead}

\begin{figure*}[t]
\centering
\includegraphics[width=0.9\textwidth]{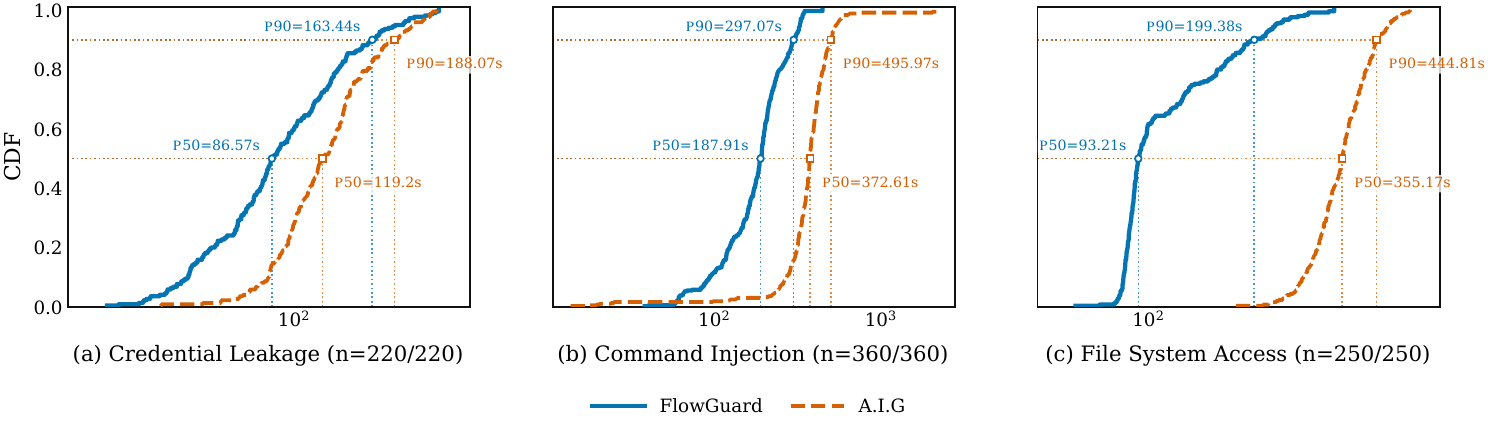}
\caption{CDF of per-case end-to-end analysis latency on three benchmark categories. The x-axis uses log scale. \ourwork's curves are consistently left-shifted, indicating lower median and tail latency than A.I.G.}
\label{fig:runtime-cdf}
\end{figure*}

This section evaluates whether \ourwork's runtime overhead is bounded and whether dynamic probing introduces unacceptable cost (RQ3).
We measure end-to-end latency per completed case, from connecting to the target MCP server to producing the final risk decision.
Figure~\ref{fig:runtime-cdf} shows the CDF of per-case latency on three representative categories.

\ourwork consistently achieves lower latency across all three categories, with left-shifted CDF curves indicating improvements in both median and tail latency. 
Compared with A.I.G, \ourwork reduces mean latency by 2.05$\times$ on Command Injection, 1.15$\times$ on Credential Leakage, and 2.23$\times$ on File System Access. 
Tail latency also decreases substantially, with P90 latency reduced from 495.97s to 297.07s, 188.07s to 163.44s, and 444.81s to 199.38s, respectively.
These results show that \ourwork improves efficiency while maintaining better detection performance (Table~\ref{tab:overall-detection}). 
The lower latency does not contradict multi-round probing: the probing budget is an upper bound, and \ourwork stops once evidence is confirmed or no new useful signal appears. As shown in Table~\ref{tab:probe-quality}, \ourwork requires fewer probes per true positive on key execution-path risks and achieves a higher evidence-call rate on Command Injection, reducing low-value probes that dominate broader dynamic scanning.
The overhead is suitable for offline scanning and CI/CD integration, and can be adapted for online deployment by separating lightweight checks from full dynamic validation.

\subsection{RQ4: Mechanism Contribution}

This section analyzes whether \ourwork's gains come from its structured verification loop rather than from using an LLM as a generic detector.
We focus on four design questions: how many probing rounds are needed, how Triage, Recon, and Feedback affect different risk types, whether response verification improves final evidence reliability, and whether the framework depends on a specific LLM backbone.

\subsubsection{Probing Budget}
We first evaluate how the probing-round budget affects \ourwork's closed-loop validation.
\ourwork does not generate a single payload and immediately make a final decision; instead, it refines later probes based on previous responses.
The probing budget therefore determines how many rounds of feedback the system can use, and directly controls the trade-off between detection effectiveness and runtime cost.

\begin{figure}[t]
\centering
\includegraphics[width=\linewidth]{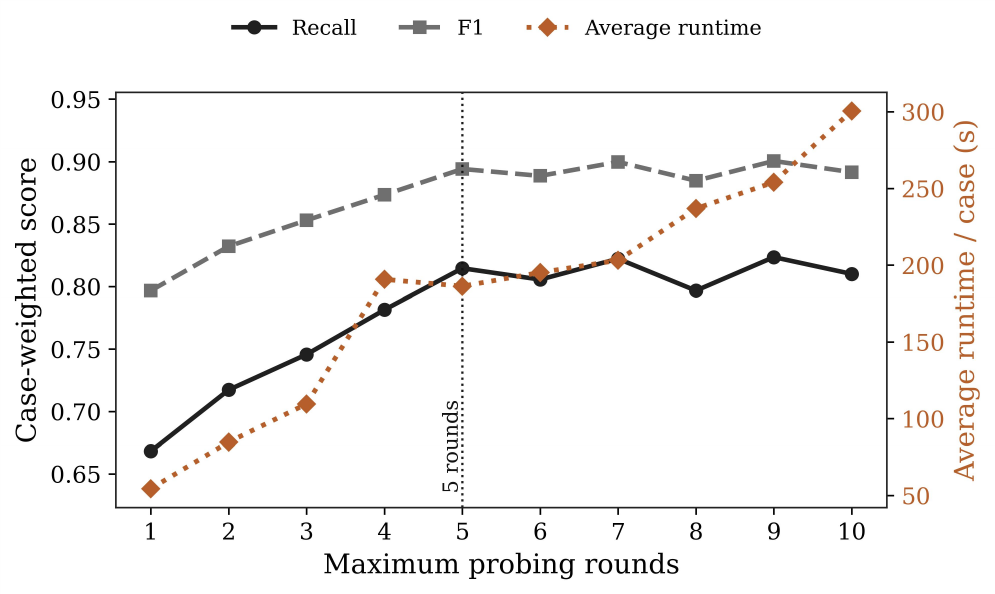}
\caption{Effect of the probing-round budget on detection effectiveness and runtime. Detection quality improves rapidly in early rounds and peaks around 5--10 rounds, while average runtime increases substantially with additional probing rounds.}
\label{fig:round_weighted}
\end{figure}

Figure~\ref{fig:round_weighted} shows Recall, F1, and mean per-case runtime across 830 cases spanning Credential Leakage, Command Injection, and File System Access. 
Detection quality improves rapidly in the early rounds. 
Recall rises from 0.668 in round 1 to 0.822 in round 5, and F1 increases from 0.797 to 0.898. 
Beyond round 5, gains largely saturate: Recall fluctuates between 0.805--0.823 and F1 between 0.885--0.900 up to round 10. 
Meanwhile, average runtime per case grows steadily from 54.4\,s in round 1 to 300.4\,s in round 10. 
These results indicate that early feedback is sufficient to repair invalid probes, specialize payloads, and focus on high-yield parameters.
The feedback loop therefore acts as bounded refinement rather than open-ended fuzzing, and we use 5 rounds as the default maximum budget.

\subsubsection{Triage, Recon, and Feedback}
We further ablate concrete workflow components to understand how structured routing, fingerprint-guided probing, and history feedback affect detection effectiveness.
Table~\ref{tab:ablation-main} reports representative results on Credential Leakage, Command Injection, and File System Access,
as these categories directly exercise active parameter-level probing and evidence validation.

\begin{table}[t]
\centering
\caption{Representative ablation results for triage, recon, and feedback.}
\label{tab:ablation-main}
\resizebox{\linewidth}{!}{
\begin{tabular}{llrrrr}
\toprule
\textbf{Task} & \textbf{Variant} & \textbf{Precision} & \textbf{Recall} & \textbf{F1} & \textbf{Accuracy} \\
\midrule
\multirow{4}{*}{\shortstack{Credential\\Leakage}}
& Full & 0.9722 & 0.7778 & 0.8642 & 0.8000 \\ 
& w/o Triage & 0.9636 & 0.2944 & 0.4511 & 0.4136 \\ 
& w/o Feedback & 0.9623 & 0.5667 & 0.7133 & 0.6273 \\ 
& w/o Recon & 1.0000 & 0.5833 & 0.7368 & 0.6591 \\ 
\midrule
\multirow{4}{*}{\shortstack{Command\\Injection}}
& Full & 1.0000 & 0.7847 & 0.8794 & 0.9139 \\ 
& w/o Triage & 1.0000 & 0.9306 & 0.9640 & 0.9722 \\ 
& w/o Feedback & 1.0000 & 0.6042 & 0.7532 & 0.8417 \\ 
& w/o Recon & 1.0000 & 0.6458 & 0.7848 & 0.8583 \\ 
\midrule
\multirow{4}{*}{\shortstack{File System\\Access}}
& Full & 1.0000 & 0.8900 & 0.9418 & 0.9120 \\
& w/o Triage & 1.0000 & 0.8350 & 0.9101 & 0.8680 \\
& w/o Feedback & 1.0000 & 0.8224 & 0.9025 & 0.8579 \\
& w/o Recon & 1.0000 & 0.1500 & 0.2609 & 0.3200 \\
\bottomrule
\end{tabular}
}
\end{table}

The results show that module contributions are risk-dependent.
For Credential Leakage, Triage is the dominant factor: removing it reduces Recall from 0.7778 to 0.2944 and F1 from 0.8642 to 0.4511.
This indicates that leakage detection needs semantic focus, because secret-like signals are sparse and the probing budget can otherwise be diluted over irrelevant tools and parameters. 
Recon and Feedback also improve leakage detection by adding runtime cues and helping distinguish true leaks from reflected or look-alike content.
Command Injection shows a different trade-off.
Recon and Feedback help validate selected parameters: removing Recon or Feedback reduces F1 to 0.7848 and 0.7532, respectively.
However, removing Triage increases Recall from 0.7847 to 0.9306.
This does not mean that Triage is unnecessary; rather, it shows that the current triage policy is too conservative for execution-path vulnerabilities, where exploitable parameters may have benign or domain-specific names. 
For File System Access, all three modules contribute meaningfully. Removing Recon drastically reduces Recall and F1, while removing Triage or Feedback leads to smaller but noticeable decreases, highlighting their complementary roles in guiding and validating probes.
Thus, Triage should be viewed as a cost-control module rather than a component that always improves accuracy in isolation.
This motivates recall-preserving triage for high-impact execution risks: skip only clearly low-risk parameters while retaining lightweight probes for uncertain ones.

\subsubsection{Evidence Adjudication}

We evaluate the Evidence Adjudication mechanism introduced in Section~\ref{sec:response-analysis},
which determines whether a candidate signal constitutes valid vulnerability evidence
by checking two dimensions: (E1) source attribution and (E2) behavioral expectation.
Table~\ref{tab:verification-ablation} reports an ablation study over these two components, including variants that use only E1, only E2, or neither.

\begin{table}[t]
\centering
\caption{Response verification ablation.}
\label{tab:verification-ablation}
\begin{tabular}{llrrr}
\toprule
\textbf{Task} & \textbf{Scheme} & \textbf{Precision} & \textbf{Recall} & \textbf{F1} \\
\midrule
\multirow{4}{*}{\shortstack{Credential\\Leakage}}
& E1 + E2 & 0.9722 & 0.7778 & 0.8642 \\
& E1 Only & 0.9787 & 0.7667 & 0.8598 \\
& E2 Only & 0.9855 & 0.7556 & 0.8553 \\
& Neither & 0.9927 & 0.7556 & 0.8580 \\
\midrule
\multirow{4}{*}{\shortstack{Command\\Injection}}
& E1 + E2 & 1.0000 & 0.7847 & 0.8794 \\
& E1 Only & 0.9910 & 0.7639 & 0.8627 \\
& E2 Only & 0.9907 & 0.7361 & 0.8446 \\
& Neither & 0.8162 & 0.7708 & 0.7929 \\
\midrule
\multirow{4}{*}{\shortstack{File System\\Access}}
& E1 + E2 & 1.0000 & 0.8900 & 0.9418 \\
& E1 Only & 1.0000 & 0.8750 & 0.9333 \\
& E2 Only & 1.0000 & 0.8800 & 0.9362 \\
& Neither & 1.0000 & 0.8850 & 0.9390 \\
\bottomrule
\end{tabular}
\end{table}

Across all tasks, combining both dimensions consistently achieves the highest F1. 
The improvement primarily comes from increasing decision reliability by filtering spurious signals that would otherwise be misclassified as vulnerabilities. 
For Command Injection, applying both E1 and E2 raises F1 to 0.8794 from 0.8446 (E2 only) or 0.7929 (neither), eliminating false positives caused by command-like outputs, benign errors, and reflected inputs. 
For Credential Leakage and File System Access, the improvement is smaller but consistent. 
These tasks already provide clearer initial signals, while Evidence Adjudication mainly improves robustness by distinguishing true system-generated outputs from benign or reflected content. 
Overall, this ablation confirms the design of Evidence Adjudication: the main gain is not extracting more suspicious strings, but preventing candidate signals from becoming findings unless they pass both source and behavioral validation.

\subsubsection{LLM Backbone Sensitivity}
Since \ourwork uses bounded semantic subroutines in triage, probe generation, and evidence attribution, we evaluate whether the verification framework depends on a particular LLM backbone.
All backbones are queried with a temperature of 0.7.

\begin{figure}[t]
    \centering
    \includegraphics[width=\linewidth]{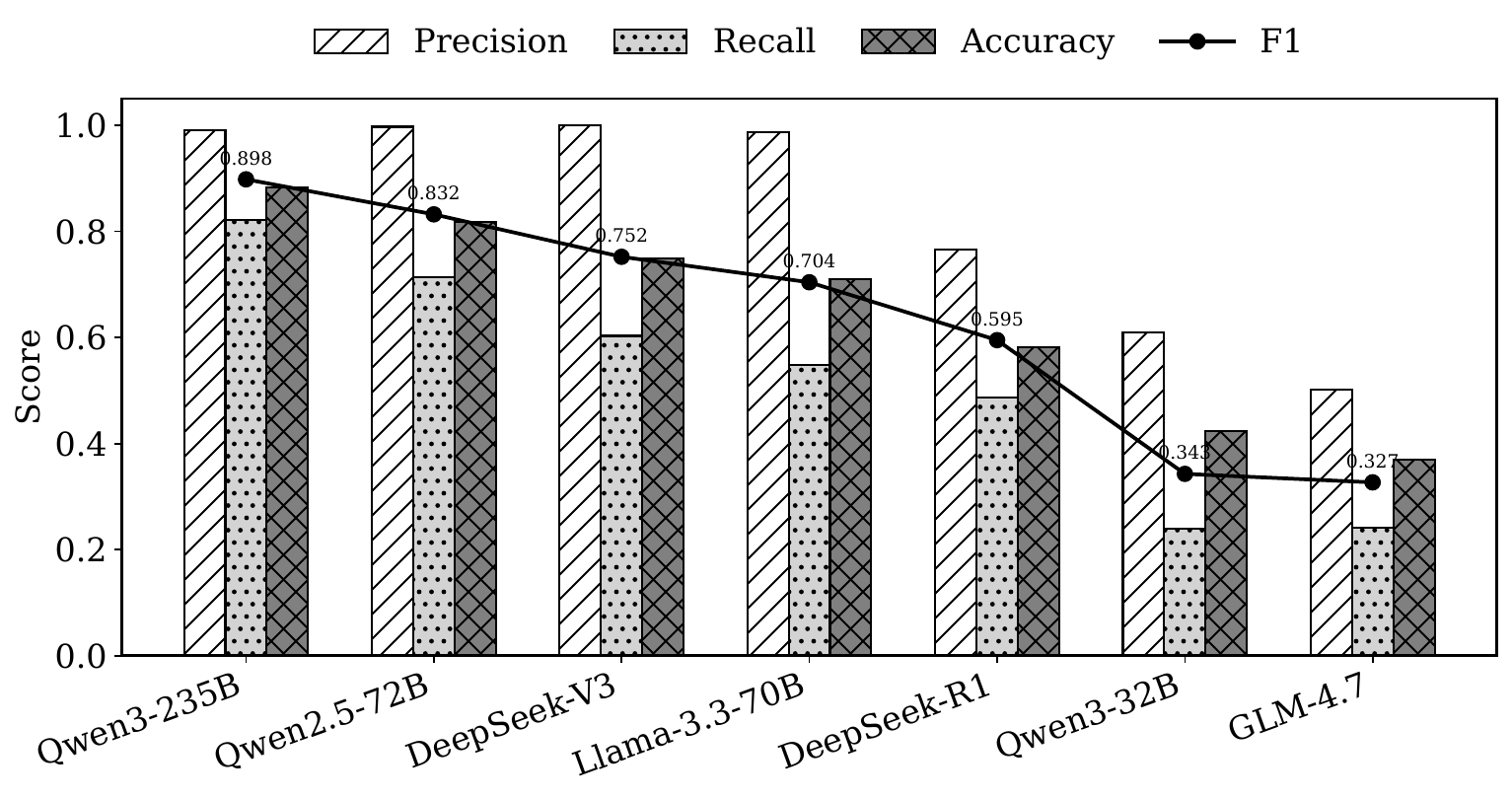}
    \caption{Sensitivity of \ourwork to different LLM backbones. Bars show Precision, Recall, and Accuracy, while the black line shows F1.}
    \label{fig:model-sensitivity}
\end{figure}

\ourwork achieves reasonable performance across a range of LLM backbones, indicating that the overall framework is not tied to a single model. 
Different models exhibit distinct trade-offs. Qwen3-235B achieves the highest overall F1 (0.898), while Qwen2.5-72B maintains similar performance with higher precision (0.997) but slightly lower recall. DeepSeek-V3 further increases precision to 1.000, at the cost of reduced recall (0.603), reflecting a more conservative detection behavior.
DeepSeek-R1 tends to be more conservative, leading to lower recall,
while Qwen3-32B and GLM-4.7 are limited by context window, resulting in missed detections.
This suggests that model capability mainly affects ambiguous semantic decisions, while the workflow structure, schema constraints, evidence checks, and stopping rules remain fixed.
Overall, these results indicate that \ourwork provides a robust verification framework across different models, and 70B-scale backbones are sufficient for practical automated assessment.
Together with the ablation results above, this shows that the performance gain comes from MCP-specific routing, schema-constrained probing, bounded feedback, and evidence adjudication rather than simply increasing fuzzing rounds or relying on a single LLM backbone.

\subsection{RQ5: Real-World Effectiveness}

\subsubsection{Server Selection}
To evaluate \ourwork in realistic MCP ecosystems, we conduct large-scale experiments on MCPZoo's runnable MCP server corpus~\cite{mcpzoo}. 
MCPZoo aggregates public servers from sources such as GitHub repositories and MCP registries, covering various domains such as developer tools, data access, cloud services, and productivity integrations. 
We select the top 8,000 runnable servers based on popularity, and interact with servers through standard MCP interfaces without independently redeploying each backend. 
We use the same server subset for \ourwork and baseline comparison.

\subsubsection{Findings}
Across the 8,000 evaluated servers, \ourwork enumerates 20,452 tools
and issues 90,079 probes.
Overall, \ourwork reports risks in 326 servers, corresponding to 4.08\% of the evaluated servers, and produces 523 individual findings.
These findings span four categories: Prompt Injection (161), Credential Leakage (155), Command Injection (129), and File System Access (78).
At the server level, Command Injection is the most widely distributed category and is reported in 115 servers.
The Command Injection, Credential Leakage, and File System Access findings report execution or disclosure evidence, whereas Prompt Injection findings represent semantic risks whose downstream impact depends on the consuming agent.
Overall, the findings span multiple MCP interaction surfaces, including returned semantic content, sensitive-information disclosure, backend command execution, and file system access.

\subsubsection{Baseline Comparison}
To examine whether \ourwork's real-world findings are also visible to existing MCP scanners, we compare against A.I.G on the same 8,000-server subset. Since MCPZoo does not provide complete vulnerability labels, this comparison characterizes reporting behavior rather than ground-truth accuracy.

On this subset, A.I.G reports risks in 4,947 servers and emits 11,055 findings, compared with \ourwork's 326 reported servers and 523 findings. The two scanners overlap on 222 reported servers, meaning A.I.G covers 68.10\% of \ourwork-reported servers, while only 4.49\% of servers reported by A.I.G are also reported by \ourwork. This gap mainly reflects different reporting criteria: A.I.G reports many metadata-level Rug Pull and Tool Poisoning risks, and 2,358 servers reported by A.I.G have no recorded MCP tool call. \ourwork requires concrete runtime evidence for execution-related findings and reports instruction-bearing outputs separately as semantic risks.

\subsubsection{Manual Verification}
To assess report quality, we manually inspect a risk-stratified random sample of 100 reported servers.
For each server, we inspect its reported findings, affected tools, probe inputs, runtime responses, risk labels, and supporting evidence generated by \ourwork.
An execution- or disclosure-related finding is considered confirmed only when the response contains concrete security-relevant evidence, such as unexpected command output, content read outside an expected filesystem boundary, or credential disclosure. Prompt Injection findings are evaluated separately as agent-dependent semantic risks because their downstream impact depends on the consuming agent and its execution policy.
A sampled server satisfies the concrete-evidence criterion if at least one of its reported findings is confirmed. In this risk-stratified sample, 84 of the 100 servers satisfy this criterion.
The remaining servers contain only agent-dependent semantic findings or findings for which the observed response is insufficient to establish exploitability.
Confirmed issues are currently undergoing responsible disclosure when appropriate. Appendix~\ref{appendix:casestudy} presents a detailed case study of a confirmed file-system vulnerability in a \texttt{convert\_to\_markdown} tool, illustrating how \ourwork's refinement mechanism surfaces non-obvious risks through iterative probe adaptation.

\section{Discussion}
\label{sec:discussion}

\subsection{From Offline Scanning to Runtime Protection}

Although \ourwork is evaluated as an offline scanner, its design naturally supports runtime deployment. 
Since MCP follows a standardized interaction workflow, \ourwork can operate as a security middleware or transparent proxy between LLM hosts and MCP servers. 
It can inspect tool metadata during Tool Discovery, analyze parameters before Tool Invocation, and examine returned content before Response Consumption. 
Depending on the detected risk, the system may log suspicious behaviors, block dangerous invocations, sanitize returned outputs, or require user confirmation.
Such deployment is particularly valuable in open MCP ecosystems, where agents frequently interact with third-party servers under different trust assumptions. 
Offline scanning and runtime guarding are complementary: offline analysis helps identify risky MCP servers before integration, while runtime monitoring can detect and mitigate unexpected behaviors that only emerge during live agent-server interaction.

\subsection{Limitations}

\ourwork has several limitations. First, as a black-box scanner, it cannot directly observe internal control flow or hidden backend logic, and may therefore miss vulnerabilities that depend on specific application states, authentication contexts, or complex multi-step interactions. 
Second, active probing introduces an inherent trade-off between coverage and operational safety. Although \ourwork uses low-side-effect probes, testing tools that interact with real files, databases, networks, or external services may still trigger unintended behaviors. 
Third, \ourwork's semantic triage, probe generation, and response attribution depend partly on the capabilities of the underlying LLM. Future advances in language models may further improve \ourwork's performance.
Finally, the current prototype focuses on non-adaptive, runtime-visible evidence from standard MCP interactions. It does not address servers that recognize the scanner and hide evidence, long-term or cross-session behavior, ecosystem-level supply-chain risks, or client-policy attacks that do not produce observable evidence within a bounded scan. Extending \ourwork to these broader risks would require more diverse client views, longer traces, and new evidence rules.

\section{Related Work}
\label{sec:related-work}

\subsection{MCP Attacks and Benchmarks}

A large body of recent work studies attacks against MCP tool metadata, tool selection, and agent behavior. Guo et al.~\cite{guo2025systematicanalysismcpsecurity} develop MCPLIB, a library of 31 MCP attack methods covering tool poisoning, credential leakage, command injection, and other MCP-specific threats. MCPTox~\cite{wang2025mcptoxbenchmarktoolpoisoning} evaluates tool poisoning attacks on real-world MCP servers. ToolHijacker~\cite{shi2026toolhijacker}, MPMA~\cite{wang2025mpmapreferencemanipulationattack}, and MCP-ITP~\cite{li2026mcpitpautomatedframeworkimplicit}, together with Mo et al.~\cite{mo2026attractivemetadataattackinducing}, show that manipulating tool names, descriptions, schemas, or poisoned metadata can bias tool retrieval, tool selection, parameter construction, and downstream tool use. Other studies examine risks beyond metadata manipulation: Maloyan and Namiot~\cite{maloyan2026breakingprotocolsecurityanalysis} and Yang et al.~\cite{yang2026compatibilitycostsystematicdiscovery} identify protocol-level and SDK-level attack opportunities, while Li et al.~\cite{li2026dontbelievereadunderstanding} study mismatches between tool descriptions and actual behavior. Beyond MCP-specific attacks, InjecAgent~\cite{zhan2024injecagent} and ToolEmu~\cite{ruan2024identifyingriskslmagents} show that indirect prompt injection and unsafe tool observations can manipulate tool-integrated agents or cause privacy leakage.
Recent MCP survey and landscape papers provide broader views of the MCP security space~\cite{hou2025mcp_landscape,li2026looksecurityissuesmodel}. \ourwork does not aim to replace these broader classifications; it focuses on risks that can be checked through observable runtime evidence.

Several benchmarks and red-teaming frameworks evaluate these risks at scale. MSB~\cite{zhang2026mcpsecuritybenchmsb} and MCPSecBench~\cite{yang2026mcpsecbenchsystematicsecuritybenchmark} evaluate agent robustness across tool discovery, invocation, and response handling, while AutoMalTool~\cite{he2025automaticredteamingllmbased} and T-MAP~\cite{lee2026tmapredteamingllmagents} automate malicious-tool generation and multi-step agent red teaming. These benchmarks mainly measure whether an agent can be induced to perform unsafe actions, usually through attack success rate or robustness metrics. 
Our benchmark instead evaluates MCP security scanners, including \ourwork. It measures whether a scanner can distinguish security risks from benign behavior in MCP interactions.

\subsection{MCP Defenses and Runtime Analysis}

Static and metadata-oriented defenses inspect MCP servers before or without executing real tool calls. MCPScan~\cite{mcpscan-antgroup} analyzes server code and configurations, while Agent-Scan~\cite{mcp_scan2025}, MCP-Scanner~\cite{mcp-scanner-cisco}, and mcp-armor~\cite{mcp-armor} audit metadata, exposed capabilities, and client settings. Semantic-artifact defenses such as MindGuard~\cite{wang2026mindguardintrinsicdecisioninspection}, MCP-GUARD~\cite{xing2026mcpguardmultistagedefenseindepthframework}, and CASCADE~\cite{turgut2026cascadecascadedhybriddefense} analyze descriptions, schemas, prompts, or returned text for poisoning and prompt-injection risks. These methods are useful for metadata-level and prompt-level threats, but they cannot verify whether a schema-valid parameter reaches a dangerous backend operation.

Runtime-oriented defenses exercise tools or analyze execution behavior. A.I.G~\cite{Tencent_AI-Infra-Guard_2025} and MCPSafetyScanner~\cite{mcpsafetyscanner} issue tool calls, while mcp-sec-audit~\cite{huang2026auditingmcpserversoverprivileged}, MCP-SandboxScan~\cite{tan2026mcpsandboxscanwasmbasedsecureexecution}, Connor~\cite{huang2026componentmanipulationcompromiseunderstanding}, and MCPShield~\cite{zhou2026mcpshieldsecuritycognitionlayer} use sandboxing, monitoring, behavioral comparison, or trust calibration. These systems show the value of runtime evidence, but they mainly audit broad capabilities or server-level behavior. \ourwork instead performs parameter-level runtime verification with schema-valid probes, reconnaissance-guided refinement, and attribution-based response adjudication.

\subsection{Fuzzing and Web Security Scanning}

Black-box fuzzing explores a target by generating inputs, observing responses, and using runtime feedback to guide subsequent tests. Snipuzz~\cite{feng2021snipuzz} infers message snippets from black-box responses, while LLM-assisted fuzzers such as ChatAFL~\cite{meng2024large} and AgentFuzz~\cite{liu2025make} use language models for protocol-state exploration, input generation, or agent-guided testing. These systems primarily optimize exploration, coverage, crash discovery, or protocol reachability. \ourwork adopts a similar feedback-guided interaction loop but applies it to evidence-grounded MCP security detection. Its objective is to determine whether tool metadata, schema-valid invocations, and returned content provide concrete security evidence rather than reflected input, benign errors, placeholders, or expected tool behavior.

Black-box web vulnerability scanners similarly discover input surfaces and validate vulnerabilities through remote interaction. BlackWidow~\cite{eriksson2021black} uses data-driven crawling and input discovery for web applications, while YuraScanner~\cite{stafeev2025yurascanner} uses LLM guidance for task-driven web scanning. MCP servers, however, expose their input surfaces through natural-language tool metadata and JSON schemas, and their responses may subsequently enter an LLM agent's context. \ourwork therefore combines schema-valid probing with response attribution to distinguish backend execution evidence from semantic content that may influence downstream agents.

\section{Conclusion}
\label{sec:conclusion}

We present \ourwork, an evidence-grounded security detection framework for MCP tools.
By combining semantic triage, recon-guided probing, and evidence adjudication, \ourwork verifies execution-related findings and identifies semantic risks without conflating them with downstream agent behavior.
On a 1,880-case executable benchmark, \ourwork achieves reliable detection effectiveness with the largest gains on execution-path vulnerabilities, while reducing end-to-end latency by up to 2.23$\times$.
These results highlight the value of separating runtime evidence from semantic risk signals in MCP security analysis.

\section*{Ethics Considerations}

This work raises three ethical considerations related to active MCP probing, real-world evaluation, and the handling of discovered security issues.

\textbf{Bounded Runtime Probing.}
\ourwork actively interacts with MCP servers to assess security risks through runtime probing. 
To reduce operational impact, the system uses bounded probe budgets, schema-valid payloads, and low-side-effect testing strategies. 
The current prototype avoids destructive actions, persistence, privilege escalation attempts, denial-of-service behaviors, or modification of external systems. 
Its goal is to identify security evidence through standard MCP interaction workflows rather than to maximize exploitation capability.

\textbf{Real-World Evaluation Scope.}
Our benchmark uses synthetic MCP servers and canary secrets rather than real credentials or private user data. 
For real-world experiments, we restrict evaluation to interactable MCP servers exposed through MCPZoo and only use standard MCP interfaces intended for normal runtime interaction. 
We do not attempt authentication bypass, infrastructure persistence, lateral movement, or access to unrelated third-party environments. 
All experiments are conducted under controlled interaction settings with bounded runtime behavior.

\textbf{Responsible Disclosure and Dual-Use Risk.}
\ourwork is designed for defensive MCP security assessment, but automated runtime probing techniques may still introduce dual-use risks if misapplied outside authorized settings. 
To reduce publication risk, the paper reports findings at the category and aggregate levels and avoids exposing real credentials, deployment details, or unnecessary high-impact payloads. 
For confirmed real-world issues, we report them to the corresponding maintainers by opening issues in the project repositories when available. The reports include the affected MCP tool and sanitized evidence, while avoiding real secrets or unnecessary exploit details.

\bibliographystyle{IEEEtran}
\bibliography{reference}
\appendices

\section{Prompt Templates}
\label{app:prompt-templates}

This appendix lists the LLM prompt locations used by \ourwork. Each prompt is used as a bounded structured subroutine in the pipeline described in Section~\ref{sec:methodology}. 

\subsection{Prompt A: LLM Triage Reasoning}
\label{app:prompt-triage}

This prompt is used in the \textsc{Triage} phase when fast-path matching cannot confidently classify a tool parameter. Its goal is to decide whether a parameter should be routed to a concrete risk family, retained for lightweight uncertainty probing, or pruned as low risk. The prompt should include the tool name, tool description, full parameter schema, target parameter name, sibling parameters, and fast-path risk scores. The expected output fields include \texttt{risk\_labels},
    \texttt{safe\_default}, \texttt{reason}, \texttt{threat\_signals}, and
    \texttt{combined\_risks}.

\begin{promptbox}{Prompt A: LLM Triage Reasoning}
You are an MCP (Model Context Protocol) tool security analyst specializing in
semantic risk triage and backend execution-path inference.

\textbf{Goal:} reduce the number of parameters that require active probing by
assigning risk labels from tool metadata, JSON Schema, parameter semantics, and
cross-parameter relationships.

Use the following criteria to produce concise JSON evidence. Do not reveal
step-by-step reasoning.

\textbf{Step 1 -- Tool Intent Recognition.}
Infer the primary backend operation from \texttt{tool\_name} and
\texttt{tool\_desc}:
\begin{itemize}
    \item File operations (read/\allowbreak{}write/\allowbreak{}delete/\allowbreak{}list/\allowbreak{}search/\allowbreak{}convert/\allowbreak{}compress)
    $\rightarrow$ prioritize FS; consider RCE if shell-like execution is suggested.
    \item Network operations (fetch/\allowbreak{}request/\allowbreak{}send/\allowbreak{}callback/\allowbreak{}webhook)
    $\rightarrow$ prioritize SSRF.
    \item Database operations (query/\allowbreak{}insert/\allowbreak{}update/\allowbreak{}delete/\allowbreak{}filter/\allowbreak{}search)
    $\rightarrow$ prioritize DB.
    \item AI/LLM operations (generate/\allowbreak{}analyze/\allowbreak{}chat/\allowbreak{}summarize)
    $\rightarrow$ prioritize PI.
    \item System operations (execute/\allowbreak{}run/\allowbreak{}schedule/\allowbreak{}config/\allowbreak{}build/\allowbreak{}deploy)
    $\rightarrow$ prioritize RCE and FS.
\end{itemize}

\textbf{Step 2 -- Per-parameter Semantic Role Inference.}
For each parameter, infer its likely backend role using concrete evidence:
\begin{itemize}
    \item Name semantics: tokens such as \texttt{path}, \texttt{file},
    \texttt{directory}, \texttt{folder}, \texttt{url}, \texttt{host},
    \texttt{query}, \texttt{command}, \texttt{script}, \texttt{prompt},
    \texttt{token}, \texttt{secret}, \texttt{key}, \texttt{config}.
    \item Description verbs: read, write, delete, list, execute, run, fetch,
    query, search, send, render, generate.
    \item Schema constraints: unconstrained strings, arrays, and objects expose
    larger input spaces; enums and booleans are usually lower risk unless their
    values select sensitive behavior.
    \item Positional role: primary operation target parameters are higher risk
    than control parameters such as \texttt{limit}, \texttt{offset},
    \texttt{sort}, \texttt{verbose}, or \texttt{dry\_run}.
\end{itemize}

\textbf{Conditional FS-to-RCE elevation.}
Assign both FS and RCE only when a path-like parameter appears in a tool whose
name or description suggests shell-like processing, external utilities, command
execution, scanning, compression, extraction, diffing, or subprocess behavior.
Otherwise, assign FS only and mention any RCE concern as conditional evidence.

\textbf{Step 3 -- Cross-parameter Combination Risk Analysis.}
Identify parameter pairs or groups whose combination creates additional risk:
\begin{itemize}
    \item \texttt{base\_dir} + \texttt{filename/path} $\rightarrow$ FS path traversal.
    \item \texttt{command/script} + \texttt{args/options} $\rightarrow$ RCE.
    \item \texttt{table/query/filter/where/search} $\rightarrow$ DB.
    \item \texttt{template/prompt/context} + \texttt{user\_input} $\rightarrow$ PI.
    \item \texttt{config/env/key/token/secret} selector +
    \texttt{output/log/debug} $\rightarrow$ CRED.
\end{itemize}

\textbf{Step 4 -- Risk Label Assignment.}
Assign one or more labels:
\begin{itemize}
    \item RCE: value may reach command execution, subprocess, shell, code
    interpreter, or dynamic evaluation.
    \item FS: value may reach file path construction, file reads/writes,
    directory traversal, archive extraction, or file deletion.
    \item DB: value may reach SQL construction, ORM filtering, database queries,
    or table/field selection.
    \item PI: value may be embedded into prompts, instructions, templates, or
    another LLM call.
    \item CRED: value may select, expose, or retrieve credentials, secrets,
    environment variables, configs, keys, or tokens.
    \item LOW: indirect or weak risk that requires additional conditions.
    \item SAFE: pure control logic with no sensitive sink, such as pagination,
    sorting, formatting, theme, or verbosity.
\end{itemize}

\textbf{Output Rules.}
\begin{itemize}
    \item Output ONLY valid JSON, with no markdown fences or extra text.
    \item For each parameter, include: \texttt{name}, \texttt{risk\_labels},
    \texttt{safe\_default}, \texttt{reason}, \texttt{threat\_signals}, and
    \texttt{combined\_risks}.
    \item \texttt{safe\_default} MUST pass the parameter's JSON Schema:
    string $\rightarrow$ \texttt{"test"}; integer $\rightarrow$ \texttt{1};
    number $\rightarrow$ \texttt{1.0}; boolean $\rightarrow$ \texttt{false};
    array $\rightarrow$ \texttt{[]}; enum $\rightarrow$ choose the first allowed
    value; object $\rightarrow$ minimal object satisfying required fields.
    \item \texttt{reason} MUST cite concrete words from the parameter name,
    parameter description, tool name, tool description, or schema.
    \item \texttt{threat\_signals} MUST list the concrete clues that triggered
    the assigned labels.
    \item If \texttt{threat\_signals} is empty, assign LOW or SAFE.
    \item Keep reasons concise and evidence-based.
\end{itemize}
\end{promptbox}

\subsection{Prompt B: Schema-Valid Probe Generation}
\label{app:prompt-probe-generation}

This prompt is used in the \textsc{Strike} phase to generate Tool Invocation arguments that preserve attack intent while satisfying the target tool's JSON schema. The prompt should include the tool metadata, target parameter schema, selected risk family, environment fingerprints from \textsc{Recon}.

\begin{promptbox}{Prompt B: Schema-Valid Probe Generation}
You are an MCP security probe generator for the Strike phase of a
Recon-then-Strike workflow.

You will receive an environment fingerprint, one high-risk parameter, its
schema constraints, and a threat family. Generate a single non-destructive
payload value for an authorized benchmark or pre-deployment security assessment.
The payload should be most likely to produce an observable security signal while
remaining schema-compatible and minimizing side effects.

\textbf{Method:}
\begin{enumerate}
    \item Respect explicit schema constraints, including type, enum, pattern,
    minimum and maximum length, numeric range, and format.
    \item Specialize the payload to the observed operating system, backend
    language, database, and threat family. Prefer compact, high-signal probes
    over generic enumeration.
    \item For FS and CRED risks, prefer benchmark-defined canary files,
    synthetic secrets, environment markers, application configuration markers,
    or other non-destructive runtime indicators when compatible with the schema.
    \item For DB and PI risks, choose a compact, constraint-aware payload
    that is likely to produce an observable signal in the response.
    \item Avoid destructive or intrusive behavior. Do not generate payloads that
    delete files, modify persistent state, exfiltrate real credentials, establish
    persistence, or contact external hosts.
\end{enumerate}

\textbf{Output Rules:}
Output only the raw payload value. Do not output JSON, markdown, explanations,
labels, or quotes unless quotes are part of the payload itself.
\end{promptbox}

\subsection{Prompt C: Evidence Adjudication}
\label{app:prompt-evidence-adjudication}

This prompt is used in the \textsc{Analysis} phase after candidate response signals have been extracted. Its goal is to decide whether the observed signal is input reflection, normal defensive behavior, or credible vulnerability evidence. The prompt should include the tool metadata, invocation arguments, raw MCP response, extracted candidate signals, expected tool semantics, and probe history. The expected output fields include \texttt{signal\_type}, \texttt{source\_attribution}, \texttt{behavioral\_expectation}, \texttt{evidence\_label}, \texttt{confidence}, and \texttt{rationale}.

\begin{promptbox}{Prompt C: Evidence Adjudication}
You are a security verdict specialist for MCP tool testing.

Your task is to determine whether a tool response constitutes genuine runtime
evidence of a vulnerability, or a false positive caused by the probe itself.
First identify whether the response contains security-relevant evidence, such as
system-originated sensitive content, execution-bearing output, file-system
feedback, or instruction-bearing content. Do not return a positive verdict
merely because the probe contains malicious-looking text. A normal error that
reflects a quoted or escaped payload as a filename or argument is not a genuine
finding.

We use the following verdict rule:
\[
R = S \land O \land \neg E.
\]
Here, \(R\) denotes \texttt{is\_real\_finding}, \(S\) denotes
\texttt{signal\_identified}, \(O\) denotes \texttt{is\_system\_originated},
and \(E\) denotes \texttt{is\_expected\_behavior}.

\textbf{Dimension 1 -- Source Attribution.}
Determine whether the sensitive or execution-bearing content was independently
produced by the tool backend, or whether it originated from the probe parameters.

Set \texttt{is\_system\_originated = false} when the response reproduces the
probe value verbatim, contains only a minor transformation of the probe value,
or when the suspicious content is fully explained by the submitted probe.
This case is treated as echo reflection.

Set \texttt{is\_system\_originated = true} when the response contains content
that was not present in the probe and appears to originate from the backend,
such as file contents, database output, environment-derived values, runtime
execution output, or backend-generated instructions.

\textbf{Dimension 2 -- Behavioral Expectation.}
Determine whether the response is expected under the tool's declared behavior.

Set \texttt{is\_expected\_behavior = true} when the tool correctly rejects the
input, returns an authentication or permission failure, reports an input
validation error, or treats the payload as a literal quoted or escaped value.
Such cases are expected defensive behavior and should not be reported as
vulnerabilities.

Set \texttt{is\_expected\_behavior = false} when the tool executes or processes
input that should have been rejected, returns content outside the declared tool
purpose, reveals internal information not implied by the tool description, or
otherwise violates its expected security contract.

\textbf{Attribution Types.}
\begin{itemize}
    \item \texttt{echo\_reflection}: suspicious content originates from the probe.
    \item \texttt{expected\_rejection}: backend-originated response, but the tool
    correctly rejected or safely handled the input.
    \item \texttt{genuine\_evidence}: backend-originated content that violates
    the expected behavior.
    \item \texttt{anomaly}: insufficient information to determine attribution.
\end{itemize}

\textbf{Content Classes.}
Classify the response content as:
\texttt{hint\_only}, \texttt{metadata\_only},
\texttt{probable\_sensitive\_value}, \texttt{confirmed\_sensitive\_value},
or \texttt{unknown}.

\textbf{Output Rules.}
Output only the following JSON object, with no markdown or extra text:
\begin{verbatim}
{
  "is_system_originated": true/false,
  "is_expected_behavior": true/false,
  "is_real_finding": true/false,
  "confidence": 0.0-1.0,
  "attribution_type": [
    "echo_reflection",
    "expected_rejection",
    "genuine_evidence",
    "anomaly"
  ],
  "content_class": [
    "hint_only",
    "metadata_only",
    "probable_sensitive_value",
    "confirmed_sensitive_value",
    "unknown"
  ],
  "reason": "one sentence overall summary"
}
\end{verbatim}
\end{promptbox}

\subsection{Prompt D: History-Guided Refinement}
\label{app:prompt-refinement}

This prompt is used in the \textsc{Refinement} phase to decide whether \ourwork should continue probing, stop without a finding, or report confirmed evidence. The prompt should include the current tool--parameter--risk tuple, environment fingerprints, previous probes, response summaries, rejected-candidate reasons, latest adjudication result, and remaining budget. 

\begin{promptbox}{Prompt D: History-Guided Refinement}
You are an expert MCP security probe planner for authorized benchmark or
pre-deployment security testing.

You are performing history-guided iterative refinement for one high-risk
parameter. You will receive a chronological probe history, schema constraints,
and runtime hints extracted from prior responses.

Your job is not to blindly mutate payloads. Your job is to:
\begin{enumerate}
    \item Infer how the target is currently interpreting the parameter at runtime.
    \item Extract actionable clues from the history, such as accepted structure,
    revealed paths, file names, table names, column names, environment-variable
    names, protocol hints, coercion hints, and failure modes.
    \item Decide the single most informative next probe for the current
    hypothesis while minimizing side effects.
\end{enumerate}

\textbf{Planning rules.}
\begin{itemize}
    \item If the history indicates schema, type, format, or protocol validation
    failure, repair the payload so it preserves the test intent while better
    satisfying the explicit constraints.
    \item If the history shows deeper runtime reachability, such as
    file-not-found, permission-denied, SQL syntax error, connection-refused,
    unsupported scheme, or target-specific parsing behavior, treat that as
    evidence about the sink and adapt the next probe to that sink.
    \item If a response reveals concrete artifacts such as file names,
    directories, table names, columns, environment-variable names, or URL
    targets, use them to plan the next probe instead of switching to an
    unrelated generic payload.
    \item If the response suggests type coercion or serialization, for example
    lists or objects being stringified before use, reason about the runtime sink
    semantics rather than only the declared schema semantics.
    \item If the history is repetitive, provides no new signal, or further
    progress is unlikely, stop.
    \item Avoid destructive actions, persistent state changes, real credential
    exfiltration, and external network contact.
\end{itemize}

\textbf{Output requirements.}
Output only valid JSON using exactly this schema:
\begin{verbatim}
{
  "decision": "continue|stop",
  "payload_mode": "declared|underlying",
  "runtime_hypothesis": "<short sentence>",
  "key_observations": ["<obs1>", "<obs2>"],
  "next_step": "<short sentence>",
  "payload": "<raw payload value or null>"
}
\end{verbatim}

If \texttt{decision} is \texttt{stop}, set \texttt{payload} to null.
Use \texttt{declared} when the next payload should still follow the declared
schema shape directly.
Use \texttt{underlying} only when the history strongly suggests the parameter is
being coerced or serialized before the sink uses it.
If \texttt{decision} is \texttt{continue}, \texttt{payload} must be the single
best next value for the target parameter, not a list.
Do not repeat a logically equivalent prior payload.
\end{promptbox}

\section{Risk Knowledge Base}

Table~\ref{tab:risk_knowledge_base} shows the risk knowledge base for Section~\ref{sec:probe-generation}.

\begin{table*}[t]
\centering
\caption{Fast-path parameter patterns used in semantic triage.}
\label{tab:risk_knowledge_base}
\resizebox{\textwidth}{!}{
\begin{tabular}{l c c l}
\toprule
\textbf{Parameter Pattern} & \textbf{Dominant Params} & \textbf{Fast-path}  & \textbf{Fast-path Parameters} \\
 & & \textbf{Params}   &  \\
\midrule
Command-Like Input & 42 & 10  &
\texttt{code}, \texttt{cmd}, \texttt{command}, \texttt{exec}, \texttt{run}, \texttt{execute}, \texttt{process}, \texttt{func}, \texttt{module}, \texttt{payload} \\

File-Like Input & 32 & 6  &
\texttt{file}, \texttt{filename}, \texttt{root}, \texttt{include}, \texttt{path}, \texttt{dir} \\

Prompt-Bearing Input & 8 & 8  &
\texttt{prompt}, \texttt{system}, \texttt{instruction}, \texttt{context}, \texttt{role}, \texttt{persona}, \texttt{memory}, \texttt{history} \\

\midrule
Total & 82 & 24 & -- \\
\bottomrule
\end{tabular}
}
\end{table*}

\section{\ourwork Verification Algorithms}
\label{app:flowguard-algorithms}

This appendix gives pseudocode for the closed-loop verification process described in Section~\ref{sec:system-overview}. Algorithm~\ref{alg:flowguard-loop} shows the full \textsc{Triage}--\textsc{Recon}--\textsc{Strike}--\textsc{Analysis}--\textsc{Refinement} loop. 

\begin{algorithm}[t]
\scriptsize
\caption{\ourwork Closed-Loop Runtime Evidence Verification}
\label{alg:flowguard-loop}
\begin{algorithmic}[1]
\Require MCP endpoint \(E\), probing budget \(B\), no-new-signal patience \(k\), risk knowledge base \(K\)
\Ensure Verified findings \(F\), inconclusive signals \(U\), scan log \(L\)
\State \(M \gets \textsc{DiscoverTools}(E)\)
\State \(C \gets \emptyset\), \(F \gets \emptyset\), \(U \gets \emptyset\), \(L \gets \emptyset\)
\ForAll{\(t \in M, p \in \textsc{Parameters}(t)\)}
        \State \(r \gets \textsc{FastPathRiskVector}(t,p,K)\)
        \If{\(\textsc{UnambiguousHighRisk}(r)\)}
            \State \(C \gets C \cup \{(t,p,\textsc{RiskFamily}(r))\}\)
        \ElsIf{\(\textsc{UncertainOrCompositional}(t,p,r)\)}
            \State \(d \gets \textsc{LLMTriage}(t,p,r)\) \Comment{Prompt A}
            \If{\(d.\textit{routing}\in\{\textsc{HighConfidenceRoute},\textsc{UncertaintyRetain}\}\)}
                \State \(C \gets C \cup \{(t,p,d.\textit{riskFamily})\}\)
            \Else
                \State \(L \gets L \cup \{\textsc{LogPruned}(t,p,d)\}\)
            \EndIf
        \EndIf
\EndFor
\ForAll{\((t,p,g) \in C\)}
    \State \(H \gets \emptyset\), \(S \gets \emptyset\), \(stale \gets 0\)
    \State \(q_{0} \gets \textsc{BuildReconProbe}(t,p,g)\)
    \State \((ok,a_{0}) \gets \textsc{ValidateSchema}(q_{0},\textsc{Schema}(t),p)\)
    \If{\(ok\)}
        \State \(y_{0} \gets \textsc{InvokeTool}(E,t,a_{0})\)
        \State \(S \gets \textsc{ExtractFingerprints}(y_{0})\)
        \State \(H \gets H \cup \{(a_{0},y_{0},\textsc{Recon})\}\)
    \EndIf
    \For{\(i=1\) to \(B\)}
        \State \(Q \gets \textsc{GenerateStrikeCandidates}(t,p,g,S,H)\) \Comment{Prompt B}
        \State \((a,rej) \gets \textsc{FirstValidCandidate}(Q,\textsc{Schema}(t),p)\)
        \If{\(a=\bot\)}
            \State \(H \gets H \cup \{(Q,rej,\textsc{Invalid})\}\)
            \State \textbf{break}
        \EndIf
        \State \(y \gets \textsc{InvokeTool}(E,t,a)\)
        \State \(Z \gets \textsc{ExtractSignals}(y,g)\)
        \State \(e \gets \textsc{AdjudicateEvidence}(t,a,y,Z,H)\) \Comment{Prompt C}
        \State \(H \gets H \cup \{(a,y,e)\}\), \(L \gets L \cup \{(t,p,g,a,e)\}\)
        \If{\(e.\textit{label}=\textsc{ConfirmedEvidence}\)}
            \State \(F \gets F \cup \{\textsc{BuildFinding}(t,p,g,a,y,e)\}\)
            \State \textbf{break}
        \EndIf
        \State \(stale \gets \textsc{UpdateStaleness}(e,H,stale)\)
        \State \(\rho \gets \textsc{PlanRefinement}(t,p,g,S,H,B-i)\) \Comment{Prompt D}
        \If{\(stale \ge k\) \textbf{ or } \(\textsc{RepeatedInvalidProbes}(H)\)}
            \State \(U \gets U \cup \{\textsc{SummarizeInconclusive}(t,p,g,H)\}\)
            \State \textbf{break}
        \ElsIf{\(\rho.\textit{decision} \in \{\textsc{Stop},\textsc{Report}\}\)}
            \State \(\textsc{ApplyRefinementDecision}(\rho,F,U,L)\)
            \State \textbf{break}
        \EndIf
    \EndFor
\EndFor
\State \Return \(F,U,L\)
\end{algorithmic}
\end{algorithm}

\vspace{-3pt}
\balance
\section{Real-World Case Study}
\label{appendix:casestudy}
This case study illustrates how \ourwork detects a non-obvious file-system vulnerability in a real-world MCP server through recon-guided refinement. 
The target is the MCP interface of Microsoft MarkItDown~\cite{markitdown}, a popular document-conversion project with over 120K GitHub stars. 
MarkItDown is designed to convert files and office documents into Markdown for downstream LLM and text-analysis workflows, making it a natural fit for MCP-based agent integrations.

\subsection{Target and Tool}
The server exposes a single tool, \texttt{convert\_to\_markdown}, which is described as converting a URI-addressed resource into Markdown format. 
The tool accepts \texttt{http:}, \texttt{https:}, \texttt{file:}, and \texttt{data:} URIs through a single \texttt{uri} parameter. 
At the metadata level, this interface appears benign: the tool name refers to document conversion, the description does not mention shell execution or file reading, and the parameter is expressed as a generic resource identifier rather than an explicit path argument. 
As a result, name- or description-based heuristics may treat the tool as a normal converter and fail to recognize that allowing caller-controlled \texttt{file:} URIs creates an attack surface for local file access. 
The risk only becomes evident when the server's runtime behavior is tested: the backend accepts a local-file URI and returns the corresponding file content.
We reported this issue through Microsoft's official MarkItDown issue tracker and received a public reply proposing that \texttt{file://} URIs be blocked by default and enabled only through an explicit \texttt{--allow-local-files} startup flag.

\subsection{Detection Process}
\ourwork first identifies the \texttt{uri} parameter as a potential file-system risk because the schema and description allow resource locations to be supplied by the caller. 
During recon, \ourwork sends a traversal-style probe, \texttt{../../../../etc/passwd}, which the server rejects because the value does not conform to a valid URI scheme. 
Rather than treating this rejection as a dead end, the refinement loop interprets it as a schema constraint signal and generates a syntactically valid follow-up probe:
\begin{verbatim}
{"uri": "file:///etc/passwd"}
\end{verbatim}
The server then returns real \texttt{/etc/passwd} content, beginning with system account records. 
The evidence adjudication step confirms that the response is system-originated rather than reflected from the probe, and that exposing local system files falls outside the intended document-conversion boundary. 
\ourwork therefore reports a confirmed file-disclosure finding with confidence 0.95.

\subsection{Root Cause}
The root cause is that the server passes the caller-controlled \texttt{uri} value to a fetching backend without enforcing a policy boundary between remote documents and local files. 
Supporting \texttt{file:} URIs may be reasonable in a trusted local CLI setting, but in an MCP deployment the same parameter becomes agent-callable. 
As a result, the converter can be repurposed into an unintended local-file disclosure primitive.

\subsection{Significance}
This case demonstrates two advantages of \ourwork. 
First, semantic triage can surface non-obvious file-system risks even when the tool metadata appears benign and lacks explicit danger keywords. 
The vulnerability is not exposed by the tool name or description alone; it arises from the combination of a caller-controlled URI parameter, support for the \texttt{file:} scheme, and the backend's failure to enforce a policy boundary between remote documents and local files. 
Second, recon-guided refinement turns an initially rejected probe into confirmed evidence by adapting to runtime schema feedback instead of abandoning the hypothesis. 
The finding was independently confirmed during manual verification.

\end{document}